\documentclass[%
 reprint,
nofootinbib,
 amsmath,amssymb,
 aps,
nolongbibliography
]{revtex4-2}

\usepackage{graphicx}

\usepackage{dcolumn}
\usepackage{bm}
\usepackage{physics}
\usepackage{placeins}
\usepackage{xcolor}
\usepackage[colorlinks=true,
            linkcolor=blue,
            citecolor=blue,
            urlcolor=blue]{hyperref}

\begin{document}

\preprint{APS/123-QED}

\title{High-Fidelity Universal Quantum Gates in a Cryogenic Atom Array}

\author{Junlan Jin${^{1,2,*}}$, Donghyuk Seo${^{1,*}}$, Yue Shi${^1}$, Yukai Lu${^1}$, Jeff D. Thompson${^2}$, Waseem S. Bakr${^{1,\dagger}}$}

\affiliation{${^1}$Department of Physics, Princeton University, Princeton, New Jersey 08544, USA\\${^2}$Department of Electrical and Computer Engineering, Princeton University, Princeton, New Jersey 08544, USA}

\date{\today}

\begin{abstract}

Neutral atoms in optical tweezer arrays have emerged as a leading platform for quantum information processing. The fidelity of entangling gates mediated by Rydberg interactions has improved rapidly and is increasingly constrained by blackbody-radiation-induced transitions, which shorten Rydberg lifetimes and create contaminant states that can cause correlated errors. Cryogenic environments can suppress these transition rates by more than an order of magnitude. However, realizing a universal gate set in such an environment has remained an outstanding challenge.
Here, we demonstrate a universal quantum gate set
in a cryogenic $^{133}$Cs atom array, with a single-qubit gate
fidelity of 99.970(2)\% and a raw two-qubit entangling-gate
fidelity of 99.76(8)\%, within the range of leading
room-temperature implementations.
The entangling gate uses direct single-photon coupling
between the electronic ground state and a Rydberg $P$ state, avoiding scattering and light shifts associated with two-photon coupling.
The gate operates at a modest Rabi frequency of $2\pi\times4.3$\,MHz.
The measured fidelity agrees with our error model dominated by atomic motion, indicating a straightforward route to higher fidelity through improved cooling, while the fundamental error contribution from Rydberg decay is below $3\times10^{-4}$. These results establish a path to entangling-gate fidelities substantially beyond the current state of the art through cryogenic suppression of Rydberg decay, reducing the overhead for fault-tolerant computation.

\end{abstract}

\maketitle
\def\thefootnote{*}\footnotetext{These authors contributed equally to this work}
\def\thefootnote{$\dagger$}\footnotetext{Email: wbakr@princeton.edu}

%%%%%%%%%%%%%%%%%%%%%%%%%%
%%%%%%%%%%%%%%%
\section{Introduction}

Neutral atoms in optical tweezer arrays have emerged as a promising platform for scalable quantum computation, combining large qubit registers, reconfigurable nonlocal connectivity, and high-fidelity quantum control~\cite{coherent-transport_bluvstein2022,99.5-fidelity_Evered2023,6100tweezer_manetsch2025}. Recent experiments have demonstrated logical-qubit operations~\cite{logical_Bluvstein2024,atom-computing_Reichardt2024, erasure_Zhang2026, logical_Rines2025, magic_SalesRodriguez2025} and repeated syndrome extraction~\cite{atom-computing_Muniz2025,fault-tolerant_Bluvstein2026,AtomComputing2026}, marking significant progress toward fault-tolerant quantum computation. Entangling gates are a key building block, and improving their fidelity can substantially reduce logical error rates and the resource overhead of fault-tolerant quantum computation~\cite{Fowler2012, Cong2022}. Recent advances have pushed entangling-gate fidelities above 99\% across multiple neutral-atom platforms~\cite{99.7_Tsai2025,99.59_liu2026,Senoo2026,99.3_Radnaev2025,AtomComputing2025Gate,99.60_lin2026,Lib2026}, with the highest reported raw fidelity reaching 99.85\%~\cite{99.85_evered2026}.

Entangling gates in neutral-atom arrays rely on transient excitation to strongly interacting Rydberg states.
Their finite lifetimes impose a fundamental limit on gate fidelity through decay during the gate \cite{Saffman2016}.
Sustained technical advances have suppressed other sources of error, such that leading gate implementations are increasingly approaching this lifetime limit~\cite{99.7_Tsai2025,99.59_liu2026,99.85_evered2026}.
At room temperature, Rydberg lifetimes are limited by both spontaneous emission and blackbody-radiation (BBR)-induced transitions to neighboring Rydberg states, with the latter making the larger contribution~\cite{Beterov2009}. Suppressing BBR-induced transitions would therefore substantially improve two-qubit gate fidelities. It would also mitigate a second problem: these transitions populate contaminant Rydberg states whose strong, long-range dipolar interactions perturb subsequent operations, introducing errors correlated across space and time~\cite{Broadening_Goldschmidt2016,Festa2022,fault-tolerant_Bluvstein2026}. Such errors can be reduced by increasing the spatial and temporal separation between gates~\cite{fault-tolerant_Bluvstein2026, 99.85_evered2026}, but at the cost of reduced gate parallelism and a lower effective clock rate.

Cryogenic operation provides a direct approach to suppressing BBR-induced transitions by reducing the thermal photon occupation at the relevant Rydberg transition frequencies~\cite{Gallagher1979BBRTheory,Spencer1981Na,Spencer1982Na,cantat2020long}. It complements techniques that engineer the electromagnetic density of states, which have extended the lifetimes of circular Rydberg states~\cite{Hulet1985,circularRydberg_Nguyen2018,circularRydberg_Cohen2021,circular-Rydberg_Wu2023, Circular_Sr_Holzl2024,pultinevicius2026long}.
Recent experiments with cryogenic atom arrays have demonstrated trapping lifetimes of thousands of seconds~\cite{6000s_Schymik2021,3000s_Zhang2025}, defect-free arrays of more than a thousand atoms~\cite{defect-free_Lim2026}, and a more than threefold enhancement of Rydberg lifetimes with an effective BBR temperature of approximately 10\,K~\cite{lifetime_Jin2026}. Despite these advances, high-fidelity universal quantum gates have yet to be demonstrated in cryogenic atom arrays. Achieving these operations requires precise coherent control of ground-state qubits and Rydberg interactions despite restricted optical access and the challenges of electric- and magnetic-field control inside the cryostat.

\begin{figure*}[t!]
    \centering
    \includegraphics[width = 1\linewidth]{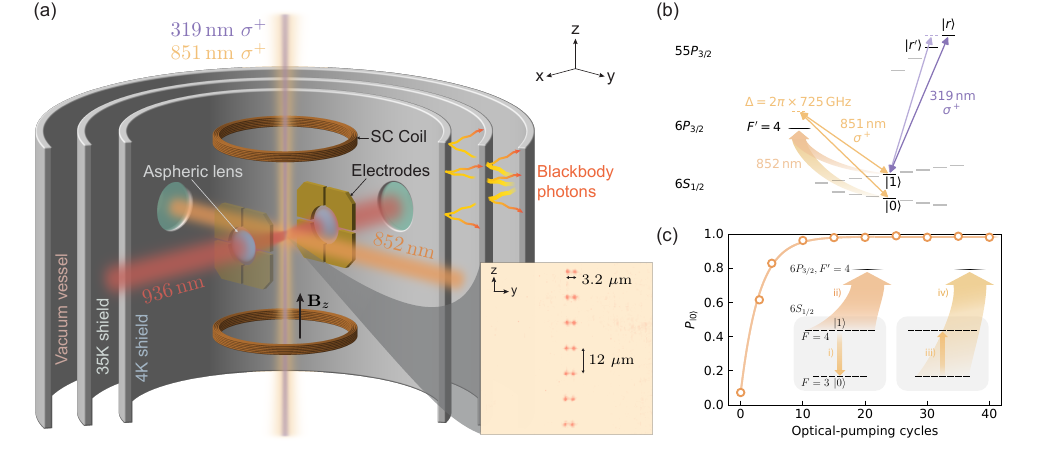}
    \caption{Experimental setup and state preparation. (a) Simplified schematic of the experimental setup, showing the 4\,K and 35\,K radiation shields enclosed within a vacuum chamber, superconducting coils, electrodes, aspheric lenses, and laser beam paths. Inset: a $7\times2$ atom array, with interpair and intrapair spacings of 12\,$\mu$m and 3.2\,$\mu$m, respectively, is located at the focus of the lenses. The tweezer trapping light is at the cesium magic wavelength of 936\,nm, where the 6$S_{1/2}$ and 6$P_{3/2}$ states have similar dynamic polarizabilities. (b) Energy-level diagram of ${}^{133}$Cs showing the relevant transitions. The qubit is encoded in the hyperfine clock states $\ket{0}=\ket{3,0}$ and $\ket{1}=\ket{4,0}$. The Raman transition is driven by far-detuned light (851\,nm), while optical pumping and repumping use near-resonant light on the cesium D2 line (852\,nm). Single-photon Rydberg excitation with $\sigma^+$-polarized light at 319\,nm couples $\ket{1}$ to $\ket{r}=\ket{55P_{3/2},m_J=3/2}$ and off-resonantly to $\ket{r'}=\ket{55P_{3/2},m_J=1/2}$. (c) Cyclic clock-state preparation. Inset: Sequence for a single preparation cycle, consisting of Raman transfers in steps i) and iii), optical pumping in step ii), and repumping in step iv).}
    \label{fig:setup_and_op}
\end{figure*}

Here, we demonstrate high-fidelity universal quantum gates in a cryogenic atom array for the first time, achieving a single-qubit gate fidelity of 99.970(2)\% and a raw two-qubit controlled-$Z$ (CZ) gate fidelity of 99.76(8)\%.
The CZ gate fidelity is comparable to leading room-temperature implementations \cite{99.7_Tsai2025,99.59_liu2026,99.60_lin2026,99.85_evered2026} despite operating at a modest excitation Rabi frequency of $\Omega=2\pi\times4.3\,$MHz.
The entangling gate employs direct single-photon excitation from the electronic ground state to a Rydberg $P$ state~\cite{dressing_Jau2016,Dressing_Zeiher2016,Schleier-Smith2023DressingSqueezing}, avoiding the short-lived intermediate state involved in conventional two-photon excitation schemes for alkali atoms.
Compared with the Rydberg $S$ states commonly used for entangling gates, $P$ states present additional challenges owing to their larger static electric polarizabilities and strongly anisotropic interactions. We address these challenges through stray-field compensation and the choice of an atom-pair geometry. The gate performance is consistent with our theoretical error model, which identifies finite-temperature atomic motion, rather than Rydberg decay, as the dominant source of infidelity.
By suppressing the fundamental error from Rydberg decay, cryogenic operation opens a route to entangling-gate fidelities substantially beyond the current state of the art.

%%%%%%%%%%%%%%%%%%%%%%%%%%
%%%%%%%%%%%%%%%
\section{Experimental setup and state preparation}
Our cryogenic apparatus, described in Ref.~\cite{lifetime_Jin2026}, consists of a room-temperature vacuum chamber enclosing two radiation shields maintained at 35\,K and 4\,K [Fig.~\ref{fig:setup_and_op}(a)]. The central assembly within the 4\,K shield contains aspheric lenses with a numerical aperture of 0.5, superconducting coils for generating magnetic fields, and electrodes for controlling the local electric field.

We encode the qubit in the hyperfine clock states $\ket{0} = \ket{F=3, m_F=0}$ and $\ket{1} = \ket{F=4, m_F=0}$ [Fig.~\ref{fig:setup_and_op}(b)], reducing sensitivity to
magnetic-field fluctuations. A magnetic field of 13.8\,G along the vertical $z$ direction, generated by the superconducting coils, defines the quantization axis and produces a Zeeman splitting of 25.8\,MHz between adjacent magnetic sublevels of the $55P_{3/2}$ Rydberg state. The field remains on throughout state preparation and gate operations and is turned off before readout.

%%%%%%%%%%%%%%%%%%%%%%%%%%
%%%%%%%%%%%%%%%
After loading, we prepare the atoms in state $\ket{1}$.
Conventional optical pumping exploits the forbidden transition $\ket{F,m_F=0}\rightarrow\ket{F'=F,m_F'=0}$ to make the target state dark to $\pi$-polarized light \cite{OP_Avila1987, 6100tweezer_manetsch2025}. Maintaining the required high-purity linear polarization at the atoms, however, is challenging because of the multiple optical windows inside the cryostat. Furthermore, the forbidden transition becomes weakly allowed in a finite magnetic field due to mixing between excited-state hyperfine levels.
Instead, we employ a cyclic optical-pumping scheme using the Cs D2 line ($6S_{1/2}\rightarrow6P_{3/2}$), similar to Ref.~\cite{op_McGuirk2002}. Co-propagating, $\pi$-polarized optical pumping and repumping beams drive the $F=4\rightarrow F'=4$ and $F=3\rightarrow F'=4$ transitions, respectively, and are applied alternately [Fig.~\ref{fig:setup_and_op}(c)]. Between optical pulses, Raman $\pi$ pulses transfer population between $\ket{4,0}$ and $\ket{3,0}$, shelving the prepared population in the hyperfine manifold not addressed by the next pulse. The shelved population is thus kept dark by the 9.2\,GHz hyperfine splitting rather than by a forbidden optical transition, making the scheme less sensitive to polarization impurity and excited-state mixing. 

For state-selective readout, we apply light resonant with the $F=4\rightarrow F'=5$ transition to remove atoms in $\ket{1}$, followed by imaging that distinguishes the qubit states by atom presence ($\ket{0}$) or absence ($\ket{1}$). To characterize the preparation fidelity, we apply a Raman pulse to transfer the population from $\ket{1}$ to $\ket{0}$, leaving any population in other $F = 4$ sublevels to be removed by the state-selective pushout. With $N=25$ cycles, 
the measured population of 98.4\% corresponds to a preparation fidelity of about 99.2\% after accounting for about 0.8\% readout loss [Fig.~\ref{fig:setup_and_op}(c)].
Raman pulse errors contribute approximately 0.1\% to the preparation infidelity, while the remainder may arise from residual spectral contamination in the optical beams, which can drive unwanted scattering from shelved atoms.

\section{Single-qubit control}
\label{Sec:1q}
\begin{figure}[t]
        \centering
        \includegraphics[width=1\linewidth]{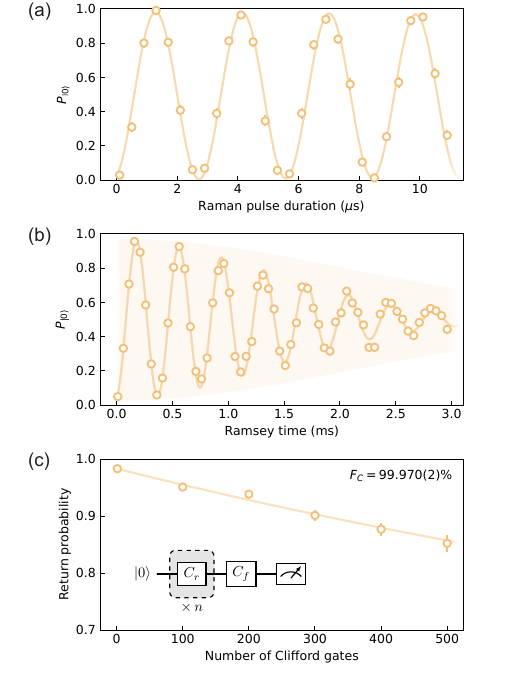}
        \caption{Coherent control of the clock-state qubit. (a) Raman-driven clock-state Rabi oscillations with a Rabi frequency of $2\pi \times 349.8(7)$\,kHz. 
        (b) Ramsey measurement. Orange points and the solid line show the array-averaged data and a Gaussian fit with a $1/e$ decay time of 1.92(4)\,ms. The shaded decay envelope corresponds to $T_2^*=3.14(2)\,$ms, the average coherence time extracted from individual-site fits.
        (c) Randomized benchmarking of single-qubit gates. An exponential fit with a baseline of $1/2$ yields an average Clifford gate fidelity of 99.970(2)\%. Inset: Randomized benchmarking circuit. $C_r$, $C_f$, and $n$ denote a random Clifford gate, the final recovery gate, and the number of Clifford gates, respectively.}
        \label{fig:1q}
\end{figure}
Having prepared the atoms in a hyperfine clock state, we next turn to single-qubit control. We drive global rotations between the clock states using a two-photon Raman transition. The Raman frequency components are generated from a single laser blue-detuned by $725\,\mathrm{GHz}$ from the $6S_{1/2}\rightarrow6P_{3/2}$ transition, using an electro-optic modulator (EOM) and a chirped Bragg grating~\cite{Raman_Levine2022}. An in-phase and quadrature (IQ)-modulated microwave drive to the EOM provides independent control of amplitude and phase, enabling arbitrary single-qubit rotations.

The Raman drive induces coherent Rabi oscillations between the clock states at a Rabi frequency of $\Omega_R=2\pi\times350\,\mathrm{kHz}$ [Fig.~\ref{fig:1q}(a)]. 
Ramsey interferometry yields a coherence time of $T_2^*=3.14(2)\,\mathrm{ms}$, obtained by averaging the coherence times extracted from fits to individual-site Ramsey signals [Fig.~\ref{fig:1q}(b)].
The observed dephasing arises mainly from variations in the tweezer-induced differential light shift caused by atomic thermal motion~\cite{dephasing_Kuhr2005,RSC_Thompson2013}, yielding a simulated coherence time of $T_2^*=4.1$\,ms at our measured temperature.
Averaging the populations across the array before fitting results in a lower value of 1.92(4)\,ms due to tweezer intensity inhomogeneity.
Applying repeated XY16 dynamical decoupling sequences~\cite{XY16_Gullion1990}, with 1\,ms between successive $\pi$ pulses, extends the array-averaged coherence time to 413(24)\,ms.

To suppress rotation errors arising from temporal drifts and spatial inhomogeneity in the Raman Rabi frequency, we employ BB1 composite pulses~\cite{BB1_Wimperis1994,NMR_Vandersypen2005}. Together with virtual $Z$ rotations~\cite{virtualZ_McKay2017}, implemented by updating the phase of subsequent Raman pulses, we realize arbitrary single-qubit rotations. The relative phase accumulated between pulses due to Raman-induced light shift is also compensated (Appendix~\ref{appendix:1q}).

We characterize the single-qubit gate fidelity using randomized benchmarking (RB)~\cite{1qrb_Knill2008, 1qrb_Magesan2011}, with each Clifford gate decomposed into physical $X$ rotations and virtual $Z$ rotations. Fitting the decay of the averaged return probability yields an average Clifford fidelity of 99.970(2)\% [Fig.~\ref{fig:1q}(c)]. We estimate that spontaneous photon scattering contributes approximately $8\times10^{-5}$ to the gate infidelity, while the remaining infidelity may arise from imperfect compensation of the differential light shift from Raman light and imperfections in the IQ modulation.

%%%%%%%%%%%%%%%%%%%%%%%%%%
%%%%%%%%%%%%%%%
\section{Two-qubit entangling gate}

Our CZ gate departs from previous implementations in two respects: it uses direct single-photon excitation from the ground state, and it is mediated by Rydberg $P$ state interactions.

%%%%%%%%%%%%%%%%%%%%%%%%%%
\subsection{Single-photon Rydberg excitation}

\begin{figure}[t]
    \centering
    \includegraphics[width=1\linewidth]{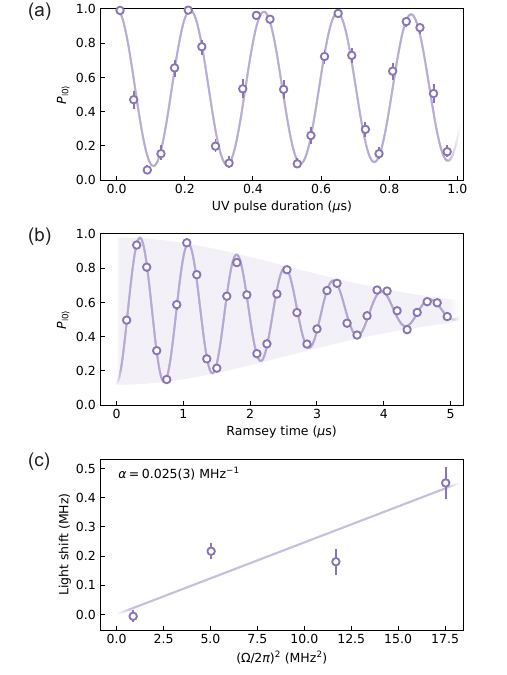}
    \caption{ Coherent control of the ground-Rydberg transition. 
    (a) Rydberg Rabi oscillations with a Rabi frequency of $2\pi \times 4.629(8)$\,MHz.
    (b) Ramsey measurement. Purple points and the solid line show the array-averaged data and a Gaussian fit with a $1/e$ decay time of $3.4(1)\,\mu$s. The shaded decay envelope corresponds to $T_2^*=3.7(1)\,\mu$s, the average coherence time extracted from individual-site fits.
    (c) Light shift of the ground-Rydberg transition.}
    \label{fig:2q_1}
\end{figure}

We employ direct single-photon excitation from $\ket{1}$ to the Rydberg state $\ket{r}=\ket{55P_{3/2},m_J=3/2}$ using $\sigma^+$-polarized 319\,nm light. Unlike conventional two-photon excitation schemes for alkali atoms, this approach avoids scattering and light shifts associated with two-photon excitation.

Our ultraviolet (UV) laser system \cite{lifetime_Jin2026} produces approximately 1\,W of light at 319\,nm, of which about 270\,mW is available at the fiber output for delivery to atoms, after losses from the acousto-optic modulator (AOM) and fiber coupling. The UV beam propagates along the quantization axis. We optimize its polarization purity by minimizing the Rabi frequency of the $\ket{1}\rightarrow\ket{55P_{3/2},m_J=-3/2}$ transition driven by residual $\sigma^-$ polarization, reducing the polarization impurity to below 0.8\% in power. With a beam waist of 25\,$\mu$m ($1/e^2$ intensity radius), we observe coherent Rydberg Rabi oscillations at a frequency of $\Omega=2\pi\times4.63$\,MHz [Fig.~\ref{fig:2q_1}(a)]. We operate at $\Omega=2\pi\times4.3\,$MHz for the entangling gates. 

We next characterize the coherence of the ground-Rydberg transition using Ramsey interferometry. Fitting the Ramsey signals at individual array sites yields an average Gaussian $1/e$ coherence time of $T_2^*=3.7(1)\,\mu$s [Fig.~\ref{fig:2q_1}(b)]. Assuming Doppler-dominated dephasing, this corresponds to a temperature of $6.0(3)\,\mu$K along the excitation beam.

We also measure the intensity-dependent light shift of the ground-Rydberg transition by tracking its resonance frequency at different Rabi frequencies. The measured shift is fitted to $\Delta\nu=\alpha(\Omega/2\pi)^2$, yielding $\alpha=0.025(3)\,\mathrm{MHz}^{-1}$ [Fig.~\ref{fig:2q_1}(c)].
This corresponds to a light shift of 0.46\,MHz at our gate operating Rabi frequency, about 40 times smaller than that
for two-photon excitation at a similar Rabi frequency~\cite{99.5-fidelity_Evered2023}. Therefore, the gate-error contribution from intensity-induced light-shift fluctuations is negligible (Appendix~\ref{Appendix:cz_err}).

\begin{figure*}[t]
    \centering
    \includegraphics[width=1\linewidth]{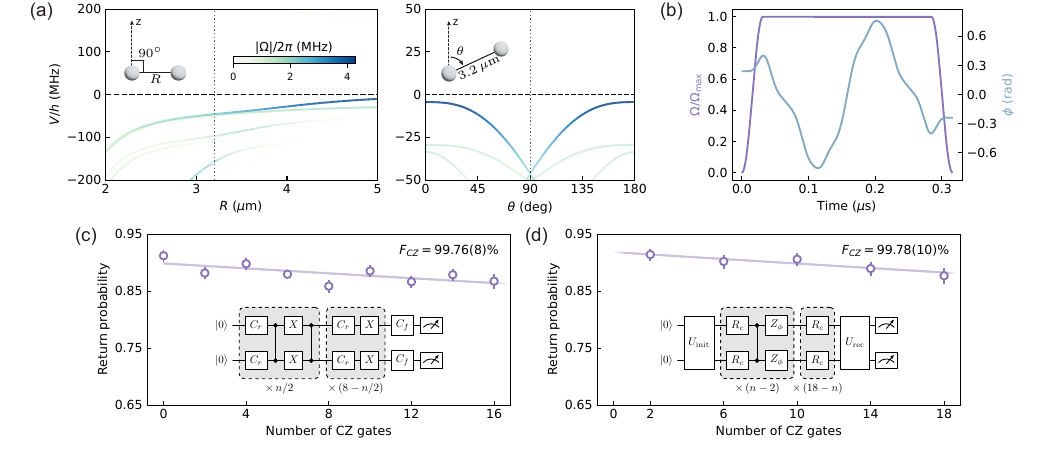}
     \caption{High-fidelity time-optimal CZ gate using Rydberg $P$ state interactions. (a) Interaction energies of relevant Rydberg pair states versus interatomic separation $R$ at $\theta=90^\circ$ (left) and versus $\theta$ at the experimental separation of $R=3.2\,\mu$m (right), calculated using ARC \cite{arc_sibalic2017}. The colors of the solid lines indicate the coupling strength between $\ket{1r}$ and the pair states. 
     (b) Amplitude (purple) and phase (light blue) profiles of the time-optimal CZ gate.
     (c) Echo randomized benchmarking and (d) symmetric stabilizer benchmarking of the CZ gate. Exponential fits with zero offset yield fidelities of 99.76(8)\% and 99.78(10)\%, respectively. Insets: Circuit diagrams. $n$ denotes the number of CZ gates. $R_c$ is a single-qubit rotation randomly drawn from $\{X_{\pm\pi/2},Y_{\pm\pi/2}\}$. $U_\mathrm{init}$ and $U_\mathrm{rec}$ are the initialization and recovery unitaries, with $U_\mathrm{init}$ preparing one of twelve symmetric stabilizer states~\cite{99.7_Tsai2025}.
     }
    \label{fig:2q_2}
\end{figure*}

%%%%%%%%%%%%%%%%%%%%%%%%%%
\subsection{Controlling Rydberg $P$ state interactions}
Having characterized coherent single-atom Rydberg excitation, we next consider the interactions that mediate the entangling gate.
While early Rydberg-blockade experiments employed Rydberg $D$ states \cite{blockade_Gaetan2009, blockade_Urban2009, CNOT_Isenhower2010,entanglement_Wilk2010}, recent high-fidelity Rydberg gates have predominantly relied on Rydberg $S$ states \cite{99.85_evered2026,99.3_Radnaev2025,99.5-fidelity_Evered2023,99.60_lin2026, 99.7_Tsai2025,AtomComputing2025Gate,99.59_liu2026,Senoo2026}, which offer isotropic interactions and relatively low electric polarizability.
Rydberg $P$ states have previously been used for Rydberg dressing to realize quantum gates \cite{dressing_Jau2016, dressing_Martin2021} and for quantum simulation~\cite{Dressing_Zeiher2016, borish_transverse-field_2020}. 
Their use in high-fidelity resonant Rydberg gates poses challenges due to their anisotropic interactions and large sensitivity to stray electric fields.

The interactions between Rydberg $P$ states depend strongly on the angle $\theta$ between the interatomic and quantization axes [Fig.~\ref{fig:2q_2}(a)]. At an interatomic separation of 3.2\,$\mu$m, the relevant pair-state potential closest to resonance has an interaction shift of $V/h\approx-45.9$\,MHz for $\theta=90^\circ$, compared with only $-4.4$\,MHz for $\theta=0^\circ$. To obtain strong intrapair interactions, we therefore orient each gate pair perpendicular to the quantization axis, arranging the atoms into seven pairs with an intrapair separation of 3.2\,$\mu$m and an interpair spacing of 12\,$\mu$m along the quantization axis [Fig.~\ref{fig:setup_and_op}(a)]. At this spacing, the interaction shift between atoms in neighboring pairs is approximately $-1.6\,$kHz, making interpair crosstalk negligible.

In addition to their anisotropic interactions, Rydberg $P$ states exhibit substantially larger static electric polarizabilities than Rydberg $S$ states, making their energies particularly sensitive to residual electric fields. For example, the polarizability of the $\ket{55P_{3/2},m_J = 3/2}$ state is approximately $2710\,$MHz\,cm$^2$/V$^2$, about 22 times that of the $\ket{55S_{1/2},m_J = 1/2}$ state. The surfaces of the aspheric lenses facing the atoms are coated with indium tin oxide (ITO) to reduce stray charge buildup. We also compensate the residual electric field using electrodes inside the 4\,K radiation shield, suppressing its magnitude to below 7\,mV/cm.

%%%%%%%%%%%%%%%%%%%%%%%%%%
\subsection{Benchmarking the time-optimal CZ gate}

Building on the characterized Rydberg excitation and interactions, we implement the time-optimal CZ gate protocol \cite{TO_Jandura2022}. We numerically optimize the laser phase profile for a fixed amplitude envelope [Fig.~\ref{fig:2q_2}(b)]. The optimization accounts for finite Rydberg interactions, off-resonant coupling to $\ket{r'}=\ket{55P_{3/2},m_J=1/2}$, and fluctuations in the interactions due to thermal motion (Appendix~\ref{appendix:subsec_pulse}). The resulting pulse implements a CZ gate up to single-qubit phases, with a duration of 316\,ns and a plateau Rabi frequency of $\Omega_\text{max}=2\pi\times4.3$\,MHz.

To characterize the entangling-gate fidelity, we implement the echo randomized benchmarking sequence \cite{99.5-fidelity_Evered2023}. The tweezers are switched off during each CZ gate to eliminate their light shift of the Rydberg transition.
We modulate the tweezer depth periodically throughout the benchmarking sequence and gradually vary the duty cycle to ramp the modulation in and out, which reduces motional heating \cite{erasure_Zhang2026}.
The modulation alternates 20\,$\mu$s tweezer-on and 1\,$\mu$s tweezer-off intervals, with each CZ gate placed in a tweezer-off interval. The single-qubit gates are performed entirely within the tweezer-on intervals, where the tweezer-induced differential light shift between the qubit states is constant, avoiding uncontrolled $Z$ rotations. 

Starting from $\ket{00}$, we vary the number of CZ gates while keeping both the number of single-qubit operations and the number of tweezer-off intervals fixed. The sequence ideally returns the atoms to $\ket{00}$. Fitting the return probability to an exponential decay with zero offset yields a raw CZ gate fidelity of 99.76(8)\% [Fig.~\ref{fig:2q_2}(c)]. 

We also perform symmetric stabilizer benchmarking following Ref.~\cite{99.7_Tsai2025}, which is designed to reduce sensitivity to single-qubit gate errors. Given our high single-qubit gate fidelity, we expect the effect of single-qubit errors on the inferred CZ gate error to be small.
This measurement yields 
a consistent fidelity of 99.78(10)\% [Fig.~\ref{fig:2q_2}(d)].

%%%%%%%%%%%%%%%%%%%%%%%%%%
\subsection{Error budget and projected improvements}

\begin{figure}
    \centering
    \includegraphics[width=1.0\linewidth]{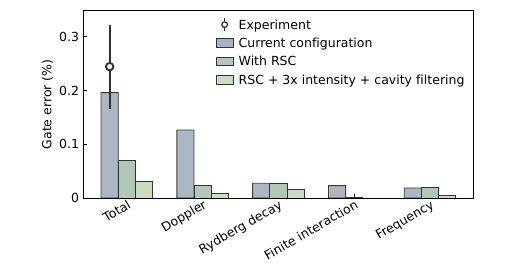}
    \caption{Simulated error budget for the time-optimal CZ gate and projected improvements. 
    Error contributions are shown for current experimental conditions, Raman sideband cooling to mean motional occupations of $\bar{n}_r=0.1$ and $\bar{n}_a=1$, and the same cooling combined with tripled excitation intensity and cavity filtering of laser frequency noise (Appendix~\ref{appendix:project_fidelity}).
    The simulated total infidelities for the three cases are $1.96\times10^{-3}$, $0.71\times10^{-3}$ and $0.31\times10^{-3}$, respectively.
    The open circle indicates the measured CZ gate error of 0.24(8)\%.
    }
    \label{fig:2q_error}
\end{figure}

Our measured CZ gate fidelity is comparable to the state of the art \cite{99.7_Tsai2025,99.59_liu2026,99.60_lin2026,99.85_evered2026} and agrees with the fidelity of 99.80\% predicted by our error model (Fig.~\ref{fig:2q_error}). The model accounts for Doppler shifts and finite-interaction fluctuations arising from thermal atomic motion, Rydberg decay, and frequency and intensity noise of the excitation laser.

Beyond the achieved fidelity, the cryogenic platform offers a distinct route to further improvement. In state-of-the-art room-temperature implementations, the finite Rydberg lifetime is a dominant source of gate error. By contrast, the extended Rydberg lifetime of approximately 406\,$\mu$s in our cryogenic system~\cite{lifetime_Jin2026} suppresses the calculated contribution from Rydberg decay to only $2.7\times10^{-4}$. The remaining infidelity is dominated by atomic motion, primarily through Doppler shifts ($1.26\times10^{-3}$), with a smaller contribution from fluctuations in the Rydberg interaction strength. Whereas increasing excitation power has been an important strategy for reducing errors in room-temperature implementations~\cite{99.85_evered2026,99.7_Tsai2025}, the distinct error budget of our cryogenic platform makes improved atomic cooling a particularly promising route to higher gate fidelities.

In our current experiment, polarization gradient cooling yields a temperature of $6.0(3)\,\mu$K along the UV excitation beam, corresponding to a mean radial motional occupation of $\bar{n}_r=2.75(17)$ at a trapping frequency of 38.9\,kHz. Raman sideband cooling (RSC) provides an established technique for reaching substantially lower temperatures, with previous experiments achieving mean motional occupations of $\bar{n}_r<0.1$ and $\bar{n}_a<1$ in the radial and axial directions, respectively~\cite{Liu2019MolecularAssembly,Spence2022GroundState,Kaufman2012GroundState, RSC_Thompson2013}. Following RSC, atoms cooled in a stretched state can be transferred to the clock state using coherent Raman pulses, minimizing additional heating.

To quantify the potential improvement, we first consider mean motional occupations of $\bar{n}_r=0.1$ and $\bar{n}_a=1$, while keeping all other parameters unchanged. Under these conditions, our model predicts a raw CZ gate fidelity of 99.93\%.
Higher excitation power offers further improvement.
UV laser output of up to 3\,W has been demonstrated at the nearby wavelength of 317\,nm~\cite{ShawThesis}, three times our current power.
Laser frequency noise can also be reduced by cavity filtering~\cite{99.7_Tsai2025}.
With these improvements and a reoptimized pulse, our model predicts a raw CZ gate fidelity of 99.97\% (Fig.~\ref{fig:2q_error} and Appendix~\ref{appendix:project_fidelity}).
Additional gains may be achieved by reducing the intrapair separation, increasing the magnetic field, and choosing a higher-lying Rydberg state with a longer lifetime.

%%%%%%%%%%%%%%%%%%%%%%%%%%
%%%%%%%%%%%%%%%
\section{Conclusions and outlook}
In this work, we have demonstrated the first universal quantum gate set in a cryogenic atom array, with single- and two-qubit gate fidelities on par with those of leading room-temperature implementations. These results show that high-fidelity universal quantum logic can be realized in a cryogenic environment that strongly suppresses BBR-induced Rydberg transitions and extends Rydberg lifetimes.
To our knowledge, our CZ gate is also the first high-fidelity entangling gate based on single-photon Rydberg excitation in alkali atoms and the first high-fidelity entangling gate mediated by Rydberg $P$ state interactions. This implementation combines precise control of residual electric fields with an atom-array geometry that accounts for the strongly anisotropic $P$ state interactions.

By suppressing errors from finite Rydberg lifetimes, cryogenic operation leaves atomic motion as the dominant source of infidelity. Our model projects a raw CZ fidelity of 99.97\% with improved cooling, higher excitation intensity, and cavity filtering of laser frequency noise. The extended Rydberg lifetime also provides greater flexibility for implementing robust gates by reducing the decay penalty associated with longer gate durations~\cite{detuning_robust_Jia2026,detuning_robust_Bataille2026,robust_Jandura2023}. More broadly, suppressing BBR-induced contaminant Rydberg states may reduce correlated errors and relax constraints on parallel and repeated gate operations.

Looking forward, the cryogenic atom array could be scaled to substantially larger sizes. We currently use an array within an 80\,$\mu$m-diameter region, limited by the field of view of the aspheric lenses. This could accommodate a few hundred tweezers at 4\,$\mu$m spacing.
Larger arrays could be realized through aberration correction~\cite{FoV_Machu2025} or in-vacuum microscope objectives~\cite{3000s_Zhang2025,defect-free_Lim2026}, which have enabled approximately two thousand tweezers in a cryogenic array~\cite{defect-free_Lim2026}.
Another constraint is the cryostat’s cooling power. At 3\,mW per tweezer and an estimated 1\% absorption of the tweezer light, the available cooling power of about 0.4\,W at the 4\,K stage can support more than ten thousand tweezers.

\textit{Acknowledgments}--- 
We would like to thank Chenyuan Li and Yicheng Bao for helpful discussions. We thank Jingxin Deng for assistance in building the single-qubit gate setup.
This work was supported by the U.S. Department of Energy, Office of Science, National Quantum Information Science Research Centers, Co-design Center for Quantum Advantage (C2QA) under contract number DE-SC0012704, the National Science Foundation (QLCI grant OMA-2120757), the Brown Science Foundation and the David and Lucile Packard Foundation (grant no. 2016-65128). D.S. was supported by the Daishin Songchon Foundation Graduate Research Fellowship.

\textit{Competing interests}--- J.D.T. is a shareholder in Logiqal, Inc.

\appendix

%%%%%%%%%%%%%%%%%%%%
\section{Experimental sequence}
\label{appendix:seq}

Atoms are first loaded into a 3D magneto-optical trap (MOT) for 200\,ms, followed by a 40\,ms optical-molasses stage for tweezer loading. The tweezers remain on throughout both stages. The total loading time of 240\,ms yields a typical array filling fraction of approximately 55\%. Both tweezers in a pair are occupied with a probability of about 30\%. The tweezers have a waist of 1.12\,$\mu$m and a trap depth of 1.1\,mK. The tweezer light is at 936\,nm, a magic wavelength for the cesium D2 line.

Following loading, we apply a first stage of polarization gradient cooling (PGC). The tweezer trap depth is then ramped down to $300\,\mu$K and the atoms are imaged with a $30\,\text{ms}$ exposure time at a detuning of $\Delta = -15\,\Gamma$, where $\Gamma=2\pi\times5.22\,$MHz is the natural linewidth of the D2 transition. Subsequently, we apply the second stage of PGC at $\Delta=-23\,\Gamma$.
We then switch to a magnetic bias field of 13.8\,G along the $z$ direction for state preparation and gate operations. For readout, we turn off the bias field, remove atoms in $F=4$ using resonant pushout light, and image the remaining atoms.

We control the magnetic field using a pair of superconducting coils inside the 4\,K radiation shield and three pairs of coils outside the room-temperature vacuum chamber.
The superconducting coils generate the magnetic-field gradient for the 3D MOT. A current of 1.2\,A through each coil produces a gradient of 8.4\,G/cm. An H-bridge switches the coils from an anti-Helmholtz to a Helmholtz configuration, providing a bias field of 13.8\,G during state preparation and gate operations. Changes in the magnetic field have a settling time of approximately 60\,ms, limited by eddy currents in the apparatus.
The external coils provide small bias fields to improve MOT–tweezer overlap during loading and compensate for background fields during PGC and imaging.
Field changes produced by these coils take about 200\,ms to settle, so they are kept at the background-field compensation settings after loading.

%%%%%%%%%%%%%%%%%%%%
\section{Single-qubit gate calibration}
\label{appendix:1q}
We generate the 9.2\,GHz microwave signal by IQ-modulating a 4.6\,GHz oscillator and subsequently frequency doubling its output. We first characterize the IQ modulator electrically, compensating DC offsets, gain imbalances, and quadrature errors. We then refine the Raman pulse amplitudes and phases using error-amplification measurements on the atoms~\cite{1q_Sheldon2016}.

The Raman light induces a differential ac Stark shift between the clock states, causing the qubit resonance frequency to differ between illuminated and dark intervals. We determine the resonance frequency under Raman illumination and use Ramsey interferometry to measure the frequency shift and calibrate the phase accumulated between pulses. At our operating intensity, the differential Stark shift is approximately 11\,kHz. We find that additional pulse-phase corrections are needed, possibly due to switching transients and pulse distortions.

To suppress errors due to amplitude variation, we implement BB1 composite rotations. In our Clifford gate compiler, there is only one type of physical gate pulse: a BB1 composite $\pi/2$ rotation consisting of the target $\pi/2$ pulse and a corrective $\pi$--$2\pi$--$\pi$ sequence. We calibrate the individual pulses while preserving their relative timing within the Raman pulse illumination window to minimize the effects of switching transients. Residual coherent-control errors, including imperfect Stark-shift compensation and IQ calibration, may contribute to the measured single-qubit gate infidelity.

%%%%%%%%%%%%%%%%%%%%%%%%%%%%%%%%%%%
\section{CZ gate design and error sources}
\label{appendix:cz_model}

\subsection{Pulse design}
\label{appendix:subsec_pulse}

We design the time-optimal CZ gate using phase-modulated excitation with a fixed amplitude envelope, following the protocol of Ref.~\cite{TO_Jandura2022}. We use a Tukey amplitude envelope with smooth rise and fall segments of \(31.4\,\mathrm{ns}\) each and a plateau Rabi frequency of $\Omega_{\max}=2\pi\times4.3\,$MHz. We numerically optimize the phase profile for a total pulse duration of 316\,ns, including finite Rydberg interactions and off-resonant coupling to the neighboring Zeeman sublevel.

The single-atom model includes the two qubit states and two Rydberg states, $\{ \ket{0}, \ket{1}, \ket{r}, \ket{r'} \}$. The excitation light resonantly couples $\ket{1}$ to $\ket{r}$ with a Rabi frequency $\Omega(t)$, while coupling $\ket{1}$ to $\ket{r'}$ with a Rabi frequency $\Omega(t)/\sqrt{3}$ and a detuning of 25.8\,MHz. Including $\ket{r'}$ in the phase-profile optimization suppresses errors from coupling to this state. The state $\ket{0}$ is treated as uncoupled.

To model the interactions, we numerically diagonalize the Rydberg-pair Hamiltonian using ARC \cite{arc_sibalic2017} and calculate the optical couplings to the resulting eigenstates.
To account for thermal fluctuations in the interatomic separation, we optimize candidate pulses at several design separations and select the pulse with the highest fidelity averaged over the thermal position distribution. For the nominal experimental separation of $3.2\,\mu$m, the selected pulse is optimized at a design separation of $3.4\,\mu$m.

We evaluate the gate fidelity averaged over pure input states drawn from the Haar measure on the full two-qubit computational space, counting population outside this space as error. The target operation is a CZ gate up to an identical single-qubit phase on each qubit.

With the designed CZ pulse waveform, we experimentally optimize the overall amplitude and center frequency by maximizing the return probability in an echo randomized benchmarking sequence containing 16 CZ gates. For symmetric stabilizer benchmarking (SSB), we additionally calibrate the single-qubit phase using an SSB sequence containing 18 CZ gates.

\subsection{Error sources}
\label{Appendix:cz_err}

Our model predicts a total gate infidelity of approximately $1.96\times10^{-3}$, dominated by Doppler shifts, with individual contributions summarized in Table~\ref{tab:cz_error_budget}.
We use a temperature of $6\,\mu$K along the excitation-beam direction, extracted from the Rydberg Ramsey measurement. Our optical pumping and repumping beams propagate along the $y$ direction [Fig.~\ref{fig:setup_and_op}(a)]. Numerical simulations suggest that this beam geometry produces approximately $3\,\mu$K more heating along $y$ than along the other directions during optical pumping. We therefore use temperatures of $(T_x,T_y,T_z)=(6,9,6)\,\mu$K in the gate-error simulations. The tweezer axis lies along $x$, and the measured radial and axial trapping frequencies are 38.9 and 5.1\,kHz, respectively.

The finite-interaction contribution includes errors from coupling to both singly and doubly excited Rydberg states, including $\ket{r'}$.

We estimate the errors from laser frequency and intensity noise using the linear-response approach \cite{99.7_Tsai2025} and the measured noise spectra of the 638\,nm light before second-harmonic generation (SHG). The calculation accounts for frequency doubling and filtering by the SHG cavity, with a linewidth of approximately 1\,MHz. The estimated contribution from laser intensity noise, including intensity-induced light-shift fluctuations, is about $1\times10^{-5}$.

We also evaluate fluctuations in the interaction angle about $\theta=90^\circ$ due to atomic motion along the $z$ direction. The radial confinement of the tweezers limits these angular fluctuations, resulting in an additional infidelity below $1\times10^{-6}$.

\begin{table}[t]
\caption{Simulated CZ gate error budget under current and projected conditions. All infidelities are in units of $10^{-4}$. The RSC projection uses $\bar{n}_r=0.1$ and $\bar{n}_a=1$. The final column additionally includes tripled excitation intensity and cavity filtering.}
\label{tab:cz_error_budget}
\begin{ruledtabular}
\begin{tabular}{lccc}
Error source & Current & RSC & RSC + $3\times$ intensity \\
             &         &     & + cavity filtering \\
\hline
Doppler shifts & 12.6 & 2.26 & 0.76 \\
Finite interaction & 2.3 & 0.10 & 0.16 \\
Rydberg decay & 2.7 & 2.68 & 1.59 \\
Laser frequency noise & 1.9 & 1.96 & 0.45 \\
Laser intensity noise & 0.1 & 0.10 & 0.10 \\
\hline
Total & 19.6 & 7.10 & 3.06 \\
\end{tabular}
\end{ruledtabular}
\end{table}

\subsection{Projected improvements}
\label{appendix:project_fidelity}

We first consider the improvement from Raman sideband cooling alone, using mean motional occupations of $\bar{n}_r=0.1$ and $\bar{n}_a=1$.
We reoptimize the pulse at the experimental interatomic separation of $3.2\,\mu$m. With all other physical parameters unchanged, the total projected infidelity decreases to $7.1\times10^{-4}$, including a Doppler contribution of $2.26\times10^{-4}$. The projected infidelity can be further reduced to $6.0\times10^{-4}$ by adiabatically lowering the tweezer depth, balancing reduced Doppler shifts against increased fluctuations in the interatomic separation.

We then consider a threefold increase in excitation intensity, corresponding to a Rabi frequency of $2\pi\times7.45\,$MHz. We also include filtering of the external-cavity diode laser (ECDL) light through an ultralow-expansion (ULE) cavity with a linewidth of approximately 50\,kHz to suppress laser frequency noise. Reoptimizing the pulse under these conditions reduces the total projected infidelity to approximately $3.1\times10^{-4}$.
Individual error contributions for both projections are listed in Table~\ref{tab:cz_error_budget}.


\begin{thebibliography}{70}%
\makeatletter
\providecommand \@ifxundefined [1]{%
 \@ifx{#1\undefined}
}%
\providecommand \@ifnum [1]{%
 \ifnum #1\expandafter \@firstoftwo
 \else \expandafter \@secondoftwo
 \fi
}%
\providecommand \@ifx [1]{%
 \ifx #1\expandafter \@firstoftwo
 \else \expandafter \@secondoftwo
 \fi
}%
\providecommand \natexlab [1]{#1}%
\providecommand \enquote  [1]{``#1''}%
\providecommand \bibnamefont  [1]{#1}%
\providecommand \bibfnamefont [1]{#1}%
\providecommand \citenamefont [1]{#1}%
\providecommand \href@noop [0]{\@secondoftwo}%
\providecommand \href [0]{\begingroup \@sanitize@url \@href}%
\providecommand \@href[1]{\@@startlink{#1}\@@href}%
\providecommand \@@href[1]{\endgroup#1\@@endlink}%
\providecommand \@sanitize@url [0]{\catcode `\\12\catcode `\$12\catcode `\&12\catcode `\#12\catcode `\^12\catcode `\_12\catcode `\%12\relax}%
\providecommand \@@startlink[1]{}%
\providecommand \@@endlink[0]{}%
\providecommand \url  [0]{\begingroup\@sanitize@url \@url }%
\providecommand \@url [1]{\endgroup\@href {#1}{\urlprefix }}%
\providecommand \urlprefix  [0]{URL }%
\providecommand \Eprint [0]{\href }%
\providecommand \doibase [0]{https://doi.org/}%
\providecommand \selectlanguage [0]{\@gobble}%
\providecommand \bibinfo  [0]{\@secondoftwo}%
\providecommand \bibfield  [0]{\@secondoftwo}%
\providecommand \translation [1]{[#1]}%
\providecommand \BibitemOpen [0]{}%
\providecommand \bibitemStop [0]{}%
\providecommand \bibitemNoStop [0]{.\EOS\space}%
\providecommand \EOS [0]{\spacefactor3000\relax}%
\providecommand \BibitemShut  [1]{\csname bibitem#1\endcsname}%
\let\auto@bib@innerbib\@empty
%</preamble>
\bibitem [{\citenamefont {Bluvstein}\ \emph {et~al.}(2022)\citenamefont {Bluvstein}, \citenamefont {Levine}, \citenamefont {Semeghini}, \citenamefont {Wang}, \citenamefont {Ebadi}, \citenamefont {Kalinowski}, \citenamefont {Keesling}, \citenamefont {Maskara}, \citenamefont {Pichler}, \citenamefont {Greiner}, \citenamefont {Vuletić},\ and\ \citenamefont {Lukin}}]{coherent-transport_bluvstein2022}%
  \BibitemOpen
  \bibfield  {author} {\bibinfo {author} {\bibfnamefont {D.}~\bibnamefont {Bluvstein}}, \bibinfo {author} {\bibfnamefont {H.}~\bibnamefont {Levine}}, \bibinfo {author} {\bibfnamefont {G.}~\bibnamefont {Semeghini}}, \bibinfo {author} {\bibfnamefont {T.~T.}\ \bibnamefont {Wang}}, \bibinfo {author} {\bibfnamefont {S.}~\bibnamefont {Ebadi}}, \bibinfo {author} {\bibfnamefont {M.}~\bibnamefont {Kalinowski}}, \bibinfo {author} {\bibfnamefont {A.}~\bibnamefont {Keesling}}, \bibinfo {author} {\bibfnamefont {N.}~\bibnamefont {Maskara}}, \bibinfo {author} {\bibfnamefont {H.}~\bibnamefont {Pichler}}, \bibinfo {author} {\bibfnamefont {M.}~\bibnamefont {Greiner}}, \bibinfo {author} {\bibfnamefont {V.}~\bibnamefont {Vuletić}},\ and\ \bibinfo {author} {\bibfnamefont {M.~D.}\ \bibnamefont {Lukin}},\ }\href {https://doi.org/10.1038/s41586-022-04592-6} {\bibfield  {journal} {\bibinfo  {journal} {Nature}\ }\textbf {\bibinfo {volume} {604}},\ \bibinfo {pages} {451} (\bibinfo {year} {2022})}\BibitemShut {NoStop}%
\bibitem [{\citenamefont {Evered}\ \emph {et~al.}(2023)\citenamefont {Evered}, \citenamefont {Bluvstein}, \citenamefont {Kalinowski}, \citenamefont {Ebadi}, \citenamefont {Manovitz}, \citenamefont {Zhou}, \citenamefont {Li}, \citenamefont {Geim}, \citenamefont {Wang}, \citenamefont {Maskara}, \citenamefont {Levine}, \citenamefont {Semeghini}, \citenamefont {Greiner}, \citenamefont {Vuletić},\ and\ \citenamefont {Lukin}}]{99.5-fidelity_Evered2023}%
  \BibitemOpen
  \bibfield  {author} {\bibinfo {author} {\bibfnamefont {S.~J.}\ \bibnamefont {Evered}}, \bibinfo {author} {\bibfnamefont {D.}~\bibnamefont {Bluvstein}}, \bibinfo {author} {\bibfnamefont {M.}~\bibnamefont {Kalinowski}}, \bibinfo {author} {\bibfnamefont {S.}~\bibnamefont {Ebadi}}, \bibinfo {author} {\bibfnamefont {T.}~\bibnamefont {Manovitz}}, \bibinfo {author} {\bibfnamefont {H.}~\bibnamefont {Zhou}}, \bibinfo {author} {\bibfnamefont {S.~H.}\ \bibnamefont {Li}}, \bibinfo {author} {\bibfnamefont {A.~A.}\ \bibnamefont {Geim}}, \bibinfo {author} {\bibfnamefont {T.~T.}\ \bibnamefont {Wang}}, \bibinfo {author} {\bibfnamefont {N.}~\bibnamefont {Maskara}}, \bibinfo {author} {\bibfnamefont {H.}~\bibnamefont {Levine}}, \bibinfo {author} {\bibfnamefont {G.}~\bibnamefont {Semeghini}}, \bibinfo {author} {\bibfnamefont {M.}~\bibnamefont {Greiner}}, \bibinfo {author} {\bibfnamefont {V.}~\bibnamefont {Vuletić}},\ and\ \bibinfo {author} {\bibfnamefont {M.~D.}\ \bibnamefont {Lukin}},\ }\href
  {https://doi.org/10.1038/s41586-023-06481-y} {\bibfield  {journal} {\bibinfo  {journal} {Nature}\ }\textbf {\bibinfo {volume} {622}},\ \bibinfo {pages} {268} (\bibinfo {year} {2023})}\BibitemShut {NoStop}%
\bibitem [{\citenamefont {Manetsch}\ \emph {et~al.}(2025)\citenamefont {Manetsch}, \citenamefont {Nomura}, \citenamefont {Bataille}, \citenamefont {Lv}, \citenamefont {Leung},\ and\ \citenamefont {Endres}}]{6100tweezer_manetsch2025}%
  \BibitemOpen
  \bibfield  {author} {\bibinfo {author} {\bibfnamefont {H.~J.}\ \bibnamefont {Manetsch}}, \bibinfo {author} {\bibfnamefont {G.}~\bibnamefont {Nomura}}, \bibinfo {author} {\bibfnamefont {E.}~\bibnamefont {Bataille}}, \bibinfo {author} {\bibfnamefont {X.}~\bibnamefont {Lv}}, \bibinfo {author} {\bibfnamefont {K.~H.}\ \bibnamefont {Leung}},\ and\ \bibinfo {author} {\bibfnamefont {M.}~\bibnamefont {Endres}},\ }\href {https://doi.org/10.1038/s41586-025-09641-4} {\bibfield  {journal} {\bibinfo  {journal} {Nature}\ }\textbf {\bibinfo {volume} {647}},\ \bibinfo {pages} {60} (\bibinfo {year} {2025})}\BibitemShut {NoStop}%
\bibitem [{\citenamefont {Bluvstein}\ \emph {et~al.}(2024)\citenamefont {Bluvstein}, \citenamefont {Evered}, \citenamefont {Geim}, \citenamefont {Li}, \citenamefont {Zhou}, \citenamefont {Manovitz}, \citenamefont {Ebadi}, \citenamefont {Cain}, \citenamefont {Kalinowski}, \citenamefont {Hangleiter}, \citenamefont {Bonilla~Ataides}, \citenamefont {Maskara}, \citenamefont {Cong}, \citenamefont {Gao}, \citenamefont {Sales~Rodriguez}, \citenamefont {Karolyshyn}, \citenamefont {Semeghini}, \citenamefont {Gullans}, \citenamefont {Greiner}, \citenamefont {Vuletić},\ and\ \citenamefont {Lukin}}]{logical_Bluvstein2024}%
  \BibitemOpen
  \bibfield  {author} {\bibinfo {author} {\bibfnamefont {D.}~\bibnamefont {Bluvstein}}, \bibinfo {author} {\bibfnamefont {S.~J.}\ \bibnamefont {Evered}}, \bibinfo {author} {\bibfnamefont {A.~A.}\ \bibnamefont {Geim}}, \bibinfo {author} {\bibfnamefont {S.~H.}\ \bibnamefont {Li}}, \bibinfo {author} {\bibfnamefont {H.}~\bibnamefont {Zhou}}, \bibinfo {author} {\bibfnamefont {T.}~\bibnamefont {Manovitz}}, \bibinfo {author} {\bibfnamefont {S.}~\bibnamefont {Ebadi}}, \bibinfo {author} {\bibfnamefont {M.}~\bibnamefont {Cain}}, \bibinfo {author} {\bibfnamefont {M.}~\bibnamefont {Kalinowski}}, \bibinfo {author} {\bibfnamefont {D.}~\bibnamefont {Hangleiter}}, \bibinfo {author} {\bibfnamefont {J.~P.}\ \bibnamefont {Bonilla~Ataides}}, \bibinfo {author} {\bibfnamefont {N.}~\bibnamefont {Maskara}}, \bibinfo {author} {\bibfnamefont {I.}~\bibnamefont {Cong}}, \bibinfo {author} {\bibfnamefont {X.}~\bibnamefont {Gao}}, \bibinfo {author} {\bibfnamefont {P.}~\bibnamefont {Sales~Rodriguez}}, \bibinfo {author} {\bibfnamefont
  {T.}~\bibnamefont {Karolyshyn}}, \bibinfo {author} {\bibfnamefont {G.}~\bibnamefont {Semeghini}}, \bibinfo {author} {\bibfnamefont {M.~J.}\ \bibnamefont {Gullans}}, \bibinfo {author} {\bibfnamefont {M.}~\bibnamefont {Greiner}}, \bibinfo {author} {\bibfnamefont {V.}~\bibnamefont {Vuletić}},\ and\ \bibinfo {author} {\bibfnamefont {M.~D.}\ \bibnamefont {Lukin}},\ }\href {https://doi.org/10.1038/s41586-023-06927-3} {\bibfield  {journal} {\bibinfo  {journal} {Nature}\ }\textbf {\bibinfo {volume} {626}},\ \bibinfo {pages} {58} (\bibinfo {year} {2024})}\BibitemShut {NoStop}%
\bibitem [{\citenamefont {Reichardt}\ \emph {et~al.}(2025)\citenamefont {Reichardt}, \citenamefont {Paetznick}, \citenamefont {Aasen}, \citenamefont {Basov}, \citenamefont {Bello-Rivas}, \citenamefont {Bonderson}, \citenamefont {Chao}, \citenamefont {van Dam}, \citenamefont {Hastings}, \citenamefont {Mishmash}, \citenamefont {Paz}, \citenamefont {da~Silva}, \citenamefont {Sundaram}, \citenamefont {Svore}, \citenamefont {Vaschillo}, \citenamefont {Wang}, \citenamefont {Zanner}, \citenamefont {Cairncross}, \citenamefont {Chen}, \citenamefont {Crow}, \citenamefont {Kim}, \citenamefont {Kindem}, \citenamefont {King}, \citenamefont {McDonald}, \citenamefont {Norcia}, \citenamefont {Ryou}, \citenamefont {Stone}, \citenamefont {Wadleigh}, \citenamefont {Barnes}, \citenamefont {Battaglino}, \citenamefont {Bohdanowicz}, \citenamefont {Booth}, \citenamefont {Brown}, \citenamefont {Brown}, \citenamefont {Cassella}, \citenamefont {Coxe}, \citenamefont {Epstein}, \citenamefont {Feldkamp}, \citenamefont {Griger},
  \citenamefont {Halperin}, \citenamefont {Heinz}, \citenamefont {Hummel}, \citenamefont {Jaffe}, \citenamefont {Jones}, \citenamefont {Kapit}, \citenamefont {Kotru}, \citenamefont {Lauigan}, \citenamefont {Li}, \citenamefont {Marjanovic}, \citenamefont {Megidish}, \citenamefont {Meredith}, \citenamefont {Morshead}, \citenamefont {Muniz}, \citenamefont {Narayanaswami}, \citenamefont {Nishiguchi}, \citenamefont {Paule}, \citenamefont {Pawlak}, \citenamefont {Pudenz}, \citenamefont {Pérez}, \citenamefont {Simon}, \citenamefont {Smull}, \citenamefont {Stack}, \citenamefont {Urbanek}, \citenamefont {van~de Veerdonk}, \citenamefont {Vendeiro}, \citenamefont {Weverka}, \citenamefont {Wilkason}, \citenamefont {Wu}, \citenamefont {Xie}, \citenamefont {Zalys-Geller}, \citenamefont {Zhang},\ and\ \citenamefont {Bloom}}]{atom-computing_Reichardt2024}%
  \BibitemOpen
  \bibfield  {author} {\bibinfo {author} {\bibfnamefont {B.~W.}\ \bibnamefont {Reichardt}}, \bibinfo {author} {\bibfnamefont {A.}~\bibnamefont {Paetznick}}, \bibinfo {author} {\bibfnamefont {D.}~\bibnamefont {Aasen}}, \bibinfo {author} {\bibfnamefont {I.}~\bibnamefont {Basov}}, \bibinfo {author} {\bibfnamefont {J.~M.}\ \bibnamefont {Bello-Rivas}}, \bibinfo {author} {\bibfnamefont {P.}~\bibnamefont {Bonderson}}, \bibinfo {author} {\bibfnamefont {R.}~\bibnamefont {Chao}}, \bibinfo {author} {\bibfnamefont {W.}~\bibnamefont {van Dam}}, \bibinfo {author} {\bibfnamefont {M.~B.}\ \bibnamefont {Hastings}}, \bibinfo {author} {\bibfnamefont {R.~V.}\ \bibnamefont {Mishmash}}, \bibinfo {author} {\bibfnamefont {A.}~\bibnamefont {Paz}}, \bibinfo {author} {\bibfnamefont {M.~P.}\ \bibnamefont {da~Silva}}, \bibinfo {author} {\bibfnamefont {A.}~\bibnamefont {Sundaram}}, \bibinfo {author} {\bibfnamefont {K.~M.}\ \bibnamefont {Svore}}, \bibinfo {author} {\bibfnamefont {A.}~\bibnamefont {Vaschillo}}, \bibinfo {author}
  {\bibfnamefont {Z.}~\bibnamefont {Wang}}, \bibinfo {author} {\bibfnamefont {M.}~\bibnamefont {Zanner}}, \bibinfo {author} {\bibfnamefont {W.~B.}\ \bibnamefont {Cairncross}}, \bibinfo {author} {\bibfnamefont {C.-A.}\ \bibnamefont {Chen}}, \bibinfo {author} {\bibfnamefont {D.}~\bibnamefont {Crow}}, \bibinfo {author} {\bibfnamefont {H.}~\bibnamefont {Kim}}, \bibinfo {author} {\bibfnamefont {J.~M.}\ \bibnamefont {Kindem}}, \bibinfo {author} {\bibfnamefont {J.}~\bibnamefont {King}}, \bibinfo {author} {\bibfnamefont {M.}~\bibnamefont {McDonald}}, \bibinfo {author} {\bibfnamefont {M.~A.}\ \bibnamefont {Norcia}}, \bibinfo {author} {\bibfnamefont {A.}~\bibnamefont {Ryou}}, \bibinfo {author} {\bibfnamefont {M.}~\bibnamefont {Stone}}, \bibinfo {author} {\bibfnamefont {L.}~\bibnamefont {Wadleigh}}, \bibinfo {author} {\bibfnamefont {K.}~\bibnamefont {Barnes}}, \bibinfo {author} {\bibfnamefont {P.}~\bibnamefont {Battaglino}}, \bibinfo {author} {\bibfnamefont {T.~C.}\ \bibnamefont {Bohdanowicz}}, \bibinfo {author}
  {\bibfnamefont {G.}~\bibnamefont {Booth}}, \bibinfo {author} {\bibfnamefont {A.}~\bibnamefont {Brown}}, \bibinfo {author} {\bibfnamefont {M.~O.}\ \bibnamefont {Brown}}, \bibinfo {author} {\bibfnamefont {K.}~\bibnamefont {Cassella}}, \bibinfo {author} {\bibfnamefont {R.}~\bibnamefont {Coxe}}, \bibinfo {author} {\bibfnamefont {J.~M.}\ \bibnamefont {Epstein}}, \bibinfo {author} {\bibfnamefont {M.}~\bibnamefont {Feldkamp}}, \bibinfo {author} {\bibfnamefont {C.}~\bibnamefont {Griger}}, \bibinfo {author} {\bibfnamefont {E.}~\bibnamefont {Halperin}}, \bibinfo {author} {\bibfnamefont {A.}~\bibnamefont {Heinz}}, \bibinfo {author} {\bibfnamefont {F.}~\bibnamefont {Hummel}}, \bibinfo {author} {\bibfnamefont {M.}~\bibnamefont {Jaffe}}, \bibinfo {author} {\bibfnamefont {A.~M.~W.}\ \bibnamefont {Jones}}, \bibinfo {author} {\bibfnamefont {E.}~\bibnamefont {Kapit}}, \bibinfo {author} {\bibfnamefont {K.}~\bibnamefont {Kotru}}, \bibinfo {author} {\bibfnamefont {J.}~\bibnamefont {Lauigan}}, \bibinfo {author} {\bibfnamefont
  {M.}~\bibnamefont {Li}}, \bibinfo {author} {\bibfnamefont {J.}~\bibnamefont {Marjanovic}}, \bibinfo {author} {\bibfnamefont {E.}~\bibnamefont {Megidish}}, \bibinfo {author} {\bibfnamefont {M.}~\bibnamefont {Meredith}}, \bibinfo {author} {\bibfnamefont {R.}~\bibnamefont {Morshead}}, \bibinfo {author} {\bibfnamefont {J.~A.}\ \bibnamefont {Muniz}}, \bibinfo {author} {\bibfnamefont {S.}~\bibnamefont {Narayanaswami}}, \bibinfo {author} {\bibfnamefont {C.}~\bibnamefont {Nishiguchi}}, \bibinfo {author} {\bibfnamefont {T.}~\bibnamefont {Paule}}, \bibinfo {author} {\bibfnamefont {K.~A.}\ \bibnamefont {Pawlak}}, \bibinfo {author} {\bibfnamefont {K.~L.}\ \bibnamefont {Pudenz}}, \bibinfo {author} {\bibfnamefont {D.~R.}\ \bibnamefont {Pérez}}, \bibinfo {author} {\bibfnamefont {J.}~\bibnamefont {Simon}}, \bibinfo {author} {\bibfnamefont {A.}~\bibnamefont {Smull}}, \bibinfo {author} {\bibfnamefont {D.}~\bibnamefont {Stack}}, \bibinfo {author} {\bibfnamefont {M.}~\bibnamefont {Urbanek}}, \bibinfo {author} {\bibfnamefont
  {R.~J.~M.}\ \bibnamefont {van~de Veerdonk}}, \bibinfo {author} {\bibfnamefont {Z.}~\bibnamefont {Vendeiro}}, \bibinfo {author} {\bibfnamefont {R.~T.}\ \bibnamefont {Weverka}}, \bibinfo {author} {\bibfnamefont {T.}~\bibnamefont {Wilkason}}, \bibinfo {author} {\bibfnamefont {T.-Y.}\ \bibnamefont {Wu}}, \bibinfo {author} {\bibfnamefont {X.}~\bibnamefont {Xie}}, \bibinfo {author} {\bibfnamefont {E.}~\bibnamefont {Zalys-Geller}}, \bibinfo {author} {\bibfnamefont {X.}~\bibnamefont {Zhang}},\ and\ \bibinfo {author} {\bibfnamefont {B.~J.}\ \bibnamefont {Bloom}},\ }\href {https://arxiv.org/abs/2411.11822} {\bibfield  {journal} {\bibinfo  {journal} {arXiv:2411.11822}\ } (\bibinfo {year} {2025})}\BibitemShut {NoStop}%
\bibitem [{\citenamefont {Zhang}\ \emph {et~al.}(2026)\citenamefont {Zhang}, \citenamefont {Liu}, \citenamefont {Bornet}, \citenamefont {Horvath}, \citenamefont {Peng}, \citenamefont {Ma}, \citenamefont {Huang}, \citenamefont {Puri},\ and\ \citenamefont {Thompson}}]{erasure_Zhang2026}%
  \BibitemOpen
  \bibfield  {author} {\bibinfo {author} {\bibfnamefont {B.}~\bibnamefont {Zhang}}, \bibinfo {author} {\bibfnamefont {G.}~\bibnamefont {Liu}}, \bibinfo {author} {\bibfnamefont {G.}~\bibnamefont {Bornet}}, \bibinfo {author} {\bibfnamefont {S.~P.}\ \bibnamefont {Horvath}}, \bibinfo {author} {\bibfnamefont {P.}~\bibnamefont {Peng}}, \bibinfo {author} {\bibfnamefont {S.}~\bibnamefont {Ma}}, \bibinfo {author} {\bibfnamefont {S.}~\bibnamefont {Huang}}, \bibinfo {author} {\bibfnamefont {S.}~\bibnamefont {Puri}},\ and\ \bibinfo {author} {\bibfnamefont {J.~D.}\ \bibnamefont {Thompson}},\ }\href {https://doi.org/10.1038/s41567-026-03309-0} {\bibfield  {journal} {\bibinfo  {journal} {Nat. Phys.}\ }\textbf {\bibinfo {volume} {22}},\ \bibinfo {pages} {910} (\bibinfo {year} {2026})}\BibitemShut {NoStop}%
\bibitem [{\citenamefont {Rines}\ \emph {et~al.}(2025)\citenamefont {Rines}, \citenamefont {Hall}, \citenamefont {Teo}, \citenamefont {Viszlai}, \citenamefont {Cole}, \citenamefont {Mason}, \citenamefont {Barker}, \citenamefont {Bedalov}, \citenamefont {Blakely}, \citenamefont {Bothwell}, \citenamefont {Carnahan}, \citenamefont {Chong}, \citenamefont {Eubanks}, \citenamefont {Fields}, \citenamefont {Gillette}, \citenamefont {Goiporia}, \citenamefont {Gokhale}, \citenamefont {Hickman}, \citenamefont {Iliev}, \citenamefont {Jones}, \citenamefont {Jones}, \citenamefont {Kuper}, \citenamefont {Lee}, \citenamefont {Lichtman}, \citenamefont {Loeffler}, \citenamefont {Mackintosh}, \citenamefont {Majdeteimouri}, \citenamefont {Mitchell}, \citenamefont {Noel}, \citenamefont {Novakoski}, \citenamefont {Omole}, \citenamefont {Owusu-Antwi}, \citenamefont {Radnaev}, \citenamefont {Reiter}, \citenamefont {Saffman}, \citenamefont {Thotakura}, \citenamefont {Tomesh},\ and\ \citenamefont {Vinogradov}}]{logical_Rines2025}%
  \BibitemOpen
  \bibfield  {author} {\bibinfo {author} {\bibfnamefont {R.}~\bibnamefont {Rines}}, \bibinfo {author} {\bibfnamefont {B.}~\bibnamefont {Hall}}, \bibinfo {author} {\bibfnamefont {M.~H.}\ \bibnamefont {Teo}}, \bibinfo {author} {\bibfnamefont {J.}~\bibnamefont {Viszlai}}, \bibinfo {author} {\bibfnamefont {D.~C.}\ \bibnamefont {Cole}}, \bibinfo {author} {\bibfnamefont {D.}~\bibnamefont {Mason}}, \bibinfo {author} {\bibfnamefont {C.}~\bibnamefont {Barker}}, \bibinfo {author} {\bibfnamefont {M.~J.}\ \bibnamefont {Bedalov}}, \bibinfo {author} {\bibfnamefont {M.}~\bibnamefont {Blakely}}, \bibinfo {author} {\bibfnamefont {T.}~\bibnamefont {Bothwell}}, \bibinfo {author} {\bibfnamefont {C.}~\bibnamefont {Carnahan}}, \bibinfo {author} {\bibfnamefont {F.~T.}\ \bibnamefont {Chong}}, \bibinfo {author} {\bibfnamefont {S.~Y.}\ \bibnamefont {Eubanks}}, \bibinfo {author} {\bibfnamefont {B.}~\bibnamefont {Fields}}, \bibinfo {author} {\bibfnamefont {M.}~\bibnamefont {Gillette}}, \bibinfo {author} {\bibfnamefont {P.}~\bibnamefont
  {Goiporia}}, \bibinfo {author} {\bibfnamefont {P.}~\bibnamefont {Gokhale}}, \bibinfo {author} {\bibfnamefont {G.~T.}\ \bibnamefont {Hickman}}, \bibinfo {author} {\bibfnamefont {M.}~\bibnamefont {Iliev}}, \bibinfo {author} {\bibfnamefont {E.~B.}\ \bibnamefont {Jones}}, \bibinfo {author} {\bibfnamefont {R.~A.}\ \bibnamefont {Jones}}, \bibinfo {author} {\bibfnamefont {K.~W.}\ \bibnamefont {Kuper}}, \bibinfo {author} {\bibfnamefont {S.}~\bibnamefont {Lee}}, \bibinfo {author} {\bibfnamefont {M.~T.}\ \bibnamefont {Lichtman}}, \bibinfo {author} {\bibfnamefont {K.}~\bibnamefont {Loeffler}}, \bibinfo {author} {\bibfnamefont {N.}~\bibnamefont {Mackintosh}}, \bibinfo {author} {\bibfnamefont {F.}~\bibnamefont {Majdeteimouri}}, \bibinfo {author} {\bibfnamefont {P.~T.}\ \bibnamefont {Mitchell}}, \bibinfo {author} {\bibfnamefont {T.~W.}\ \bibnamefont {Noel}}, \bibinfo {author} {\bibfnamefont {E.}~\bibnamefont {Novakoski}}, \bibinfo {author} {\bibfnamefont {V.}~\bibnamefont {Omole}}, \bibinfo {author} {\bibfnamefont
  {D.}~\bibnamefont {Owusu-Antwi}}, \bibinfo {author} {\bibfnamefont {A.~G.}\ \bibnamefont {Radnaev}}, \bibinfo {author} {\bibfnamefont {A.}~\bibnamefont {Reiter}}, \bibinfo {author} {\bibfnamefont {M.}~\bibnamefont {Saffman}}, \bibinfo {author} {\bibfnamefont {B.}~\bibnamefont {Thotakura}}, \bibinfo {author} {\bibfnamefont {T.}~\bibnamefont {Tomesh}},\ and\ \bibinfo {author} {\bibfnamefont {I.}~\bibnamefont {Vinogradov}},\ }\href {https://arxiv.org/abs/2509.13247} {\bibfield  {journal} {\bibinfo  {journal} {arXiv:2509.13247}\ } (\bibinfo {year} {2025})}\BibitemShut {NoStop}%
\bibitem [{\citenamefont {Sales~Rodriguez}\ \emph {et~al.}(2025)\citenamefont {Sales~Rodriguez}, \citenamefont {Robinson}, \citenamefont {Jepsen}, \citenamefont {He}, \citenamefont {Duckering}, \citenamefont {Zhao}, \citenamefont {Wu}, \citenamefont {Campo}, \citenamefont {Bagnall}, \citenamefont {Kwon}, \citenamefont {Karolyshyn}, \citenamefont {Weinberg}, \citenamefont {Cain}, \citenamefont {Evered}, \citenamefont {Geim}, \citenamefont {Kalinowski}, \citenamefont {Li}, \citenamefont {Manovitz}, \citenamefont {Amato-Grill}, \citenamefont {Basham}, \citenamefont {Bernstein}, \citenamefont {Braverman}, \citenamefont {Bylinskii}, \citenamefont {Choukri}, \citenamefont {DeAngelo}, \citenamefont {Fang}, \citenamefont {Fieweger}, \citenamefont {Frederick}, \citenamefont {Haines}, \citenamefont {Hamdan}, \citenamefont {Hammett}, \citenamefont {Hsu}, \citenamefont {Hu}, \citenamefont {Huber}, \citenamefont {Jia}, \citenamefont {Kedar}, \citenamefont {Kornja{\v c}a}, \citenamefont {Liu}, \citenamefont {Long},
  \citenamefont {Lopatin}, \citenamefont {Lopes}, \citenamefont {Luo}, \citenamefont {Macr{\`\i}}, \citenamefont {Markovi{\'c}}, \citenamefont {Mart{\'\i}nez-Mart{\'\i}nez}, \citenamefont {Meng}, \citenamefont {Ostermann}, \citenamefont {Ostroumov}, \citenamefont {Paquette}, \citenamefont {Qiang}, \citenamefont {Shofman}, \citenamefont {Singh}, \citenamefont {Singh}, \citenamefont {Sinha}, \citenamefont {Thoreen}, \citenamefont {Wan}, \citenamefont {Wang}, \citenamefont {Waxman-Lenz}, \citenamefont {Wong}, \citenamefont {Wurtz}, \citenamefont {Zhdanov}, \citenamefont {Zheng}, \citenamefont {Greiner}, \citenamefont {Keesling}, \citenamefont {Gemelke}, \citenamefont {Vuleti{\'c}}, \citenamefont {Kitagawa}, \citenamefont {Wang}, \citenamefont {Bluvstein}, \citenamefont {Lukin}, \citenamefont {Lukin}, \citenamefont {Zhou},\ and\ \citenamefont {Cant{\'u}}}]{magic_SalesRodriguez2025}%
  \BibitemOpen
  \bibfield  {author} {\bibinfo {author} {\bibfnamefont {P.}~\bibnamefont {Sales~Rodriguez}}, \bibinfo {author} {\bibfnamefont {J.~M.}\ \bibnamefont {Robinson}}, \bibinfo {author} {\bibfnamefont {P.~N.}\ \bibnamefont {Jepsen}}, \bibinfo {author} {\bibfnamefont {Z.}~\bibnamefont {He}}, \bibinfo {author} {\bibfnamefont {C.}~\bibnamefont {Duckering}}, \bibinfo {author} {\bibfnamefont {C.}~\bibnamefont {Zhao}}, \bibinfo {author} {\bibfnamefont {K.-H.}\ \bibnamefont {Wu}}, \bibinfo {author} {\bibfnamefont {J.}~\bibnamefont {Campo}}, \bibinfo {author} {\bibfnamefont {K.}~\bibnamefont {Bagnall}}, \bibinfo {author} {\bibfnamefont {M.}~\bibnamefont {Kwon}}, \bibinfo {author} {\bibfnamefont {T.}~\bibnamefont {Karolyshyn}}, \bibinfo {author} {\bibfnamefont {P.}~\bibnamefont {Weinberg}}, \bibinfo {author} {\bibfnamefont {M.}~\bibnamefont {Cain}}, \bibinfo {author} {\bibfnamefont {S.~J.}\ \bibnamefont {Evered}}, \bibinfo {author} {\bibfnamefont {A.~A.}\ \bibnamefont {Geim}}, \bibinfo {author} {\bibfnamefont
  {M.}~\bibnamefont {Kalinowski}}, \bibinfo {author} {\bibfnamefont {S.~H.}\ \bibnamefont {Li}}, \bibinfo {author} {\bibfnamefont {T.}~\bibnamefont {Manovitz}}, \bibinfo {author} {\bibfnamefont {J.}~\bibnamefont {Amato-Grill}}, \bibinfo {author} {\bibfnamefont {J.~I.}\ \bibnamefont {Basham}}, \bibinfo {author} {\bibfnamefont {L.}~\bibnamefont {Bernstein}}, \bibinfo {author} {\bibfnamefont {B.}~\bibnamefont {Braverman}}, \bibinfo {author} {\bibfnamefont {A.}~\bibnamefont {Bylinskii}}, \bibinfo {author} {\bibfnamefont {A.}~\bibnamefont {Choukri}}, \bibinfo {author} {\bibfnamefont {R.~J.}\ \bibnamefont {DeAngelo}}, \bibinfo {author} {\bibfnamefont {F.}~\bibnamefont {Fang}}, \bibinfo {author} {\bibfnamefont {C.}~\bibnamefont {Fieweger}}, \bibinfo {author} {\bibfnamefont {P.}~\bibnamefont {Frederick}}, \bibinfo {author} {\bibfnamefont {D.}~\bibnamefont {Haines}}, \bibinfo {author} {\bibfnamefont {M.}~\bibnamefont {Hamdan}}, \bibinfo {author} {\bibfnamefont {J.}~\bibnamefont {Hammett}}, \bibinfo {author}
  {\bibfnamefont {N.}~\bibnamefont {Hsu}}, \bibinfo {author} {\bibfnamefont {M.-G.}\ \bibnamefont {Hu}}, \bibinfo {author} {\bibfnamefont {F.}~\bibnamefont {Huber}}, \bibinfo {author} {\bibfnamefont {N.}~\bibnamefont {Jia}}, \bibinfo {author} {\bibfnamefont {D.}~\bibnamefont {Kedar}}, \bibinfo {author} {\bibfnamefont {M.}~\bibnamefont {Kornja{\v c}a}}, \bibinfo {author} {\bibfnamefont {F.}~\bibnamefont {Liu}}, \bibinfo {author} {\bibfnamefont {J.}~\bibnamefont {Long}}, \bibinfo {author} {\bibfnamefont {J.}~\bibnamefont {Lopatin}}, \bibinfo {author} {\bibfnamefont {P.~L.~S.}\ \bibnamefont {Lopes}}, \bibinfo {author} {\bibfnamefont {X.-Z.}\ \bibnamefont {Luo}}, \bibinfo {author} {\bibfnamefont {T.}~\bibnamefont {Macr{\`\i}}}, \bibinfo {author} {\bibfnamefont {O.}~\bibnamefont {Markovi{\'c}}}, \bibinfo {author} {\bibfnamefont {L.~A.}\ \bibnamefont {Mart{\'\i}nez-Mart{\'\i}nez}}, \bibinfo {author} {\bibfnamefont {X.}~\bibnamefont {Meng}}, \bibinfo {author} {\bibfnamefont {S.}~\bibnamefont {Ostermann}}, \bibinfo
  {author} {\bibfnamefont {E.}~\bibnamefont {Ostroumov}}, \bibinfo {author} {\bibfnamefont {D.}~\bibnamefont {Paquette}}, \bibinfo {author} {\bibfnamefont {Z.}~\bibnamefont {Qiang}}, \bibinfo {author} {\bibfnamefont {V.}~\bibnamefont {Shofman}}, \bibinfo {author} {\bibfnamefont {A.}~\bibnamefont {Singh}}, \bibinfo {author} {\bibfnamefont {M.}~\bibnamefont {Singh}}, \bibinfo {author} {\bibfnamefont {N.}~\bibnamefont {Sinha}}, \bibinfo {author} {\bibfnamefont {H.}~\bibnamefont {Thoreen}}, \bibinfo {author} {\bibfnamefont {N.}~\bibnamefont {Wan}}, \bibinfo {author} {\bibfnamefont {Y.}~\bibnamefont {Wang}}, \bibinfo {author} {\bibfnamefont {D.}~\bibnamefont {Waxman-Lenz}}, \bibinfo {author} {\bibfnamefont {T.}~\bibnamefont {Wong}}, \bibinfo {author} {\bibfnamefont {J.}~\bibnamefont {Wurtz}}, \bibinfo {author} {\bibfnamefont {A.}~\bibnamefont {Zhdanov}}, \bibinfo {author} {\bibfnamefont {L.}~\bibnamefont {Zheng}}, \bibinfo {author} {\bibfnamefont {M.}~\bibnamefont {Greiner}}, \bibinfo {author} {\bibfnamefont
  {A.}~\bibnamefont {Keesling}}, \bibinfo {author} {\bibfnamefont {N.}~\bibnamefont {Gemelke}}, \bibinfo {author} {\bibfnamefont {V.}~\bibnamefont {Vuleti{\'c}}}, \bibinfo {author} {\bibfnamefont {T.}~\bibnamefont {Kitagawa}}, \bibinfo {author} {\bibfnamefont {S.-T.}\ \bibnamefont {Wang}}, \bibinfo {author} {\bibfnamefont {D.}~\bibnamefont {Bluvstein}}, \bibinfo {author} {\bibfnamefont {M.~D.}\ \bibnamefont {Lukin}}, \bibinfo {author} {\bibfnamefont {A.}~\bibnamefont {Lukin}}, \bibinfo {author} {\bibfnamefont {H.}~\bibnamefont {Zhou}},\ and\ \bibinfo {author} {\bibfnamefont {S.~H.}\ \bibnamefont {Cant{\'u}}},\ }\href {https://doi.org/10.1038/s41586-025-09367-3} {\bibfield  {journal} {\bibinfo  {journal} {Nature}\ }\textbf {\bibinfo {volume} {645}},\ \bibinfo {pages} {620} (\bibinfo {year} {2025})}\BibitemShut {NoStop}%
\bibitem [{\citenamefont {Muniz}\ \emph {et~al.}(2025{\natexlab{a}})\citenamefont {Muniz}, \citenamefont {Crow}, \citenamefont {Kim}, \citenamefont {Kindem}, \citenamefont {Cairncross}, \citenamefont {Ryou}, \citenamefont {Bohdanowicz}, \citenamefont {Chen}, \citenamefont {Ji}, \citenamefont {Jones}, \citenamefont {Megidish}, \citenamefont {Nishiguchi}, \citenamefont {Urbanek}, \citenamefont {Wadleigh}, \citenamefont {Wilkason}, \citenamefont {Aasen}, \citenamefont {Barnes}, \citenamefont {Bello-Rivas}, \citenamefont {Bloomfield}, \citenamefont {Booth}, \citenamefont {Brown}, \citenamefont {Brown}, \citenamefont {Cassella}, \citenamefont {Cowan}, \citenamefont {Epstein}, \citenamefont {Feldkamp}, \citenamefont {Griger}, \citenamefont {Hassan}, \citenamefont {Heinz}, \citenamefont {Halperin}, \citenamefont {Hofler}, \citenamefont {Hummel}, \citenamefont {Jaffe}, \citenamefont {Kapit}, \citenamefont {Kotru}, \citenamefont {Lauigan}, \citenamefont {Marjanovic}, \citenamefont {Meredith}, \citenamefont {McDonald},
  \citenamefont {Morshead}, \citenamefont {Narayanaswami}, \citenamefont {Pawlak}, \citenamefont {Pudenz}, \citenamefont {P\'erez}, \citenamefont {Sabharwal}, \citenamefont {Simon}, \citenamefont {Smull}, \citenamefont {Sorensen}, \citenamefont {Stack}, \citenamefont {Stone}, \citenamefont {Taneja}, \citenamefont {van~de Veerdonk}, \citenamefont {Vendeiro}, \citenamefont {Weverka}, \citenamefont {White}, \citenamefont {Wu}, \citenamefont {Xie}, \citenamefont {Zalys-Geller}, \citenamefont {Zhang}, \citenamefont {King}, \citenamefont {Bloom},\ and\ \citenamefont {Norcia}}]{atom-computing_Muniz2025}%
  \BibitemOpen
  \bibfield  {author} {\bibinfo {author} {\bibfnamefont {J.~A.}\ \bibnamefont {Muniz}}, \bibinfo {author} {\bibfnamefont {D.}~\bibnamefont {Crow}}, \bibinfo {author} {\bibfnamefont {H.}~\bibnamefont {Kim}}, \bibinfo {author} {\bibfnamefont {J.~M.}\ \bibnamefont {Kindem}}, \bibinfo {author} {\bibfnamefont {W.~B.}\ \bibnamefont {Cairncross}}, \bibinfo {author} {\bibfnamefont {A.}~\bibnamefont {Ryou}}, \bibinfo {author} {\bibfnamefont {T.~C.}\ \bibnamefont {Bohdanowicz}}, \bibinfo {author} {\bibfnamefont {C.-A.}\ \bibnamefont {Chen}}, \bibinfo {author} {\bibfnamefont {Y.}~\bibnamefont {Ji}}, \bibinfo {author} {\bibfnamefont {A.~M.~W.}\ \bibnamefont {Jones}}, \bibinfo {author} {\bibfnamefont {E.}~\bibnamefont {Megidish}}, \bibinfo {author} {\bibfnamefont {C.}~\bibnamefont {Nishiguchi}}, \bibinfo {author} {\bibfnamefont {M.}~\bibnamefont {Urbanek}}, \bibinfo {author} {\bibfnamefont {L.}~\bibnamefont {Wadleigh}}, \bibinfo {author} {\bibfnamefont {T.}~\bibnamefont {Wilkason}}, \bibinfo {author} {\bibfnamefont
  {D.}~\bibnamefont {Aasen}}, \bibinfo {author} {\bibfnamefont {K.}~\bibnamefont {Barnes}}, \bibinfo {author} {\bibfnamefont {J.~M.}\ \bibnamefont {Bello-Rivas}}, \bibinfo {author} {\bibfnamefont {I.}~\bibnamefont {Bloomfield}}, \bibinfo {author} {\bibfnamefont {G.}~\bibnamefont {Booth}}, \bibinfo {author} {\bibfnamefont {A.}~\bibnamefont {Brown}}, \bibinfo {author} {\bibfnamefont {M.~O.}\ \bibnamefont {Brown}}, \bibinfo {author} {\bibfnamefont {K.}~\bibnamefont {Cassella}}, \bibinfo {author} {\bibfnamefont {G.}~\bibnamefont {Cowan}}, \bibinfo {author} {\bibfnamefont {J.}~\bibnamefont {Epstein}}, \bibinfo {author} {\bibfnamefont {M.}~\bibnamefont {Feldkamp}}, \bibinfo {author} {\bibfnamefont {C.}~\bibnamefont {Griger}}, \bibinfo {author} {\bibfnamefont {Y.}~\bibnamefont {Hassan}}, \bibinfo {author} {\bibfnamefont {A.}~\bibnamefont {Heinz}}, \bibinfo {author} {\bibfnamefont {E.}~\bibnamefont {Halperin}}, \bibinfo {author} {\bibfnamefont {T.}~\bibnamefont {Hofler}}, \bibinfo {author} {\bibfnamefont
  {F.}~\bibnamefont {Hummel}}, \bibinfo {author} {\bibfnamefont {M.}~\bibnamefont {Jaffe}}, \bibinfo {author} {\bibfnamefont {E.}~\bibnamefont {Kapit}}, \bibinfo {author} {\bibfnamefont {K.}~\bibnamefont {Kotru}}, \bibinfo {author} {\bibfnamefont {J.}~\bibnamefont {Lauigan}}, \bibinfo {author} {\bibfnamefont {J.}~\bibnamefont {Marjanovic}}, \bibinfo {author} {\bibfnamefont {M.}~\bibnamefont {Meredith}}, \bibinfo {author} {\bibfnamefont {M.}~\bibnamefont {McDonald}}, \bibinfo {author} {\bibfnamefont {R.}~\bibnamefont {Morshead}}, \bibinfo {author} {\bibfnamefont {S.}~\bibnamefont {Narayanaswami}}, \bibinfo {author} {\bibfnamefont {K.~A.}\ \bibnamefont {Pawlak}}, \bibinfo {author} {\bibfnamefont {K.~L.}\ \bibnamefont {Pudenz}}, \bibinfo {author} {\bibfnamefont {D.~R.}\ \bibnamefont {P\'erez}}, \bibinfo {author} {\bibfnamefont {P.}~\bibnamefont {Sabharwal}}, \bibinfo {author} {\bibfnamefont {J.}~\bibnamefont {Simon}}, \bibinfo {author} {\bibfnamefont {A.}~\bibnamefont {Smull}}, \bibinfo {author} {\bibfnamefont
  {M.}~\bibnamefont {Sorensen}}, \bibinfo {author} {\bibfnamefont {D.~T.}\ \bibnamefont {Stack}}, \bibinfo {author} {\bibfnamefont {M.}~\bibnamefont {Stone}}, \bibinfo {author} {\bibfnamefont {L.}~\bibnamefont {Taneja}}, \bibinfo {author} {\bibfnamefont {R.~J.~M.}\ \bibnamefont {van~de Veerdonk}}, \bibinfo {author} {\bibfnamefont {Z.}~\bibnamefont {Vendeiro}}, \bibinfo {author} {\bibfnamefont {R.~T.}\ \bibnamefont {Weverka}}, \bibinfo {author} {\bibfnamefont {K.}~\bibnamefont {White}}, \bibinfo {author} {\bibfnamefont {T.-Y.}\ \bibnamefont {Wu}}, \bibinfo {author} {\bibfnamefont {X.}~\bibnamefont {Xie}}, \bibinfo {author} {\bibfnamefont {E.}~\bibnamefont {Zalys-Geller}}, \bibinfo {author} {\bibfnamefont {X.}~\bibnamefont {Zhang}}, \bibinfo {author} {\bibfnamefont {J.}~\bibnamefont {King}}, \bibinfo {author} {\bibfnamefont {B.~J.}\ \bibnamefont {Bloom}},\ and\ \bibinfo {author} {\bibfnamefont {M.~A.}\ \bibnamefont {Norcia}},\ }\href {https://doi.org/10.1103/v7ny-fg31} {\bibfield  {journal} {\bibinfo  {journal}
  {Phys. Rev. X}\ }\textbf {\bibinfo {volume} {15}},\ \bibinfo {pages} {041040} (\bibinfo {year} {2025}{\natexlab{a}})}\BibitemShut {NoStop}%
\bibitem [{\citenamefont {Bluvstein}\ \emph {et~al.}(2026)\citenamefont {Bluvstein}, \citenamefont {Geim}, \citenamefont {Li}, \citenamefont {Evered}, \citenamefont {Bonilla~Ataides}, \citenamefont {Baranes}, \citenamefont {Gu}, \citenamefont {Manovitz}, \citenamefont {Xu}, \citenamefont {Kalinowski}, \citenamefont {Majidy}, \citenamefont {Kokail}, \citenamefont {Maskara}, \citenamefont {Trapp}, \citenamefont {Stewart}, \citenamefont {Hollerith}, \citenamefont {Zhou}, \citenamefont {Gullans}, \citenamefont {Yelin}, \citenamefont {Greiner}, \citenamefont {Vuletić}, \citenamefont {Cain},\ and\ \citenamefont {Lukin}}]{fault-tolerant_Bluvstein2026}%
  \BibitemOpen
  \bibfield  {author} {\bibinfo {author} {\bibfnamefont {D.}~\bibnamefont {Bluvstein}}, \bibinfo {author} {\bibfnamefont {A.~A.}\ \bibnamefont {Geim}}, \bibinfo {author} {\bibfnamefont {S.~H.}\ \bibnamefont {Li}}, \bibinfo {author} {\bibfnamefont {S.~J.}\ \bibnamefont {Evered}}, \bibinfo {author} {\bibfnamefont {J.~P.}\ \bibnamefont {Bonilla~Ataides}}, \bibinfo {author} {\bibfnamefont {G.}~\bibnamefont {Baranes}}, \bibinfo {author} {\bibfnamefont {A.}~\bibnamefont {Gu}}, \bibinfo {author} {\bibfnamefont {T.}~\bibnamefont {Manovitz}}, \bibinfo {author} {\bibfnamefont {M.}~\bibnamefont {Xu}}, \bibinfo {author} {\bibfnamefont {M.}~\bibnamefont {Kalinowski}}, \bibinfo {author} {\bibfnamefont {S.}~\bibnamefont {Majidy}}, \bibinfo {author} {\bibfnamefont {C.}~\bibnamefont {Kokail}}, \bibinfo {author} {\bibfnamefont {N.}~\bibnamefont {Maskara}}, \bibinfo {author} {\bibfnamefont {E.~C.}\ \bibnamefont {Trapp}}, \bibinfo {author} {\bibfnamefont {L.~M.}\ \bibnamefont {Stewart}}, \bibinfo {author} {\bibfnamefont
  {S.}~\bibnamefont {Hollerith}}, \bibinfo {author} {\bibfnamefont {H.}~\bibnamefont {Zhou}}, \bibinfo {author} {\bibfnamefont {M.~J.}\ \bibnamefont {Gullans}}, \bibinfo {author} {\bibfnamefont {S.~F.}\ \bibnamefont {Yelin}}, \bibinfo {author} {\bibfnamefont {M.}~\bibnamefont {Greiner}}, \bibinfo {author} {\bibfnamefont {V.}~\bibnamefont {Vuletić}}, \bibinfo {author} {\bibfnamefont {M.}~\bibnamefont {Cain}},\ and\ \bibinfo {author} {\bibfnamefont {M.~D.}\ \bibnamefont {Lukin}},\ }\href {https://doi.org/10.1038/s41586-025-09848-5} {\bibfield  {journal} {\bibinfo  {journal} {Nature}\ }\textbf {\bibinfo {volume} {649}},\ \bibinfo {pages} {39} (\bibinfo {year} {2026})}\BibitemShut {NoStop}%
\bibitem [{\citenamefont {{Atom Computing and Collaborators}}(2026)}]{AtomComputing2026}%
  \BibitemOpen
  \bibfield  {author} {\bibinfo {author} {\bibnamefont {{Atom Computing and Collaborators}}},\ }\href {https://arxiv.org/abs/2606.04079} {\bibfield  {journal} {\bibinfo  {journal} {arXiv:2606.04079}\ } (\bibinfo {year} {2026})}\BibitemShut {NoStop}%
\bibitem [{\citenamefont {Fowler}\ \emph {et~al.}(2012)\citenamefont {Fowler}, \citenamefont {Mariantoni}, \citenamefont {Martinis},\ and\ \citenamefont {Cleland}}]{Fowler2012}%
  \BibitemOpen
  \bibfield  {author} {\bibinfo {author} {\bibfnamefont {A.~G.}\ \bibnamefont {Fowler}}, \bibinfo {author} {\bibfnamefont {M.}~\bibnamefont {Mariantoni}}, \bibinfo {author} {\bibfnamefont {J.~M.}\ \bibnamefont {Martinis}},\ and\ \bibinfo {author} {\bibfnamefont {A.~N.}\ \bibnamefont {Cleland}},\ }\href {https://doi.org/10.1103/PhysRevA.86.032324} {\bibfield  {journal} {\bibinfo  {journal} {Phys. Rev. A}\ }\textbf {\bibinfo {volume} {86}},\ \bibinfo {pages} {032324} (\bibinfo {year} {2012})}\BibitemShut {NoStop}%
\bibitem [{\citenamefont {Cong}\ \emph {et~al.}(2022)\citenamefont {Cong}, \citenamefont {Levine}, \citenamefont {Keesling}, \citenamefont {Bluvstein}, \citenamefont {Wang},\ and\ \citenamefont {Lukin}}]{Cong2022}%
  \BibitemOpen
  \bibfield  {author} {\bibinfo {author} {\bibfnamefont {I.}~\bibnamefont {Cong}}, \bibinfo {author} {\bibfnamefont {H.}~\bibnamefont {Levine}}, \bibinfo {author} {\bibfnamefont {A.}~\bibnamefont {Keesling}}, \bibinfo {author} {\bibfnamefont {D.}~\bibnamefont {Bluvstein}}, \bibinfo {author} {\bibfnamefont {S.-T.}\ \bibnamefont {Wang}},\ and\ \bibinfo {author} {\bibfnamefont {M.~D.}\ \bibnamefont {Lukin}},\ }\href {https://doi.org/10.1103/PhysRevX.12.021049} {\bibfield  {journal} {\bibinfo  {journal} {Phys. Rev. X}\ }\textbf {\bibinfo {volume} {12}},\ \bibinfo {pages} {021049} (\bibinfo {year} {2022})}\BibitemShut {NoStop}%
\bibitem [{\citenamefont {Tsai}\ \emph {et~al.}(2025)\citenamefont {Tsai}, \citenamefont {Sun}, \citenamefont {Shaw}, \citenamefont {Finkelstein},\ and\ \citenamefont {Endres}}]{99.7_Tsai2025}%
  \BibitemOpen
  \bibfield  {author} {\bibinfo {author} {\bibfnamefont {R.~B.-S.}\ \bibnamefont {Tsai}}, \bibinfo {author} {\bibfnamefont {X.}~\bibnamefont {Sun}}, \bibinfo {author} {\bibfnamefont {A.~L.}\ \bibnamefont {Shaw}}, \bibinfo {author} {\bibfnamefont {R.}~\bibnamefont {Finkelstein}},\ and\ \bibinfo {author} {\bibfnamefont {M.}~\bibnamefont {Endres}},\ }\href {https://doi.org/10.1103/PRXQuantum.6.010331} {\bibfield  {journal} {\bibinfo  {journal} {PRX Quantum}\ }\textbf {\bibinfo {volume} {6}},\ \bibinfo {pages} {010331} (\bibinfo {year} {2025})}\BibitemShut {NoStop}%
\bibitem [{\citenamefont {Liu}\ \emph {et~al.}(2026)\citenamefont {Liu}, \citenamefont {Bornet}, \citenamefont {Kurdak}, \citenamefont {Xiao}, \citenamefont {Li}, \citenamefont {Zhang},\ and\ \citenamefont {Thompson}}]{99.59_liu2026}%
  \BibitemOpen
  \bibfield  {author} {\bibinfo {author} {\bibfnamefont {G.}~\bibnamefont {Liu}}, \bibinfo {author} {\bibfnamefont {G.}~\bibnamefont {Bornet}}, \bibinfo {author} {\bibfnamefont {D.}~\bibnamefont {Kurdak}}, \bibinfo {author} {\bibfnamefont {M.}~\bibnamefont {Xiao}}, \bibinfo {author} {\bibfnamefont {C.}~\bibnamefont {Li}}, \bibinfo {author} {\bibfnamefont {B.}~\bibnamefont {Zhang}},\ and\ \bibinfo {author} {\bibfnamefont {J.~D.}\ \bibnamefont {Thompson}},\ }\href {https://arxiv.org/abs/2606.05060} {\bibfield  {journal} {\bibinfo  {journal} {arXiv:2606.05060}\ } (\bibinfo {year} {2026})}\BibitemShut {NoStop}%
\bibitem [{\citenamefont {Senoo}\ \emph {et~al.}(2026)\citenamefont {Senoo}, \citenamefont {Baumgärtner}, \citenamefont {Lis}, \citenamefont {Vaidya}, \citenamefont {Zeng}, \citenamefont {Giudici}, \citenamefont {Pichler},\ and\ \citenamefont {Kaufman}}]{Senoo2026}%
  \BibitemOpen
  \bibfield  {author} {\bibinfo {author} {\bibfnamefont {A.}~\bibnamefont {Senoo}}, \bibinfo {author} {\bibfnamefont {A.}~\bibnamefont {Baumgärtner}}, \bibinfo {author} {\bibfnamefont {J.~W.}\ \bibnamefont {Lis}}, \bibinfo {author} {\bibfnamefont {G.~M.}\ \bibnamefont {Vaidya}}, \bibinfo {author} {\bibfnamefont {Z.}~\bibnamefont {Zeng}}, \bibinfo {author} {\bibfnamefont {G.}~\bibnamefont {Giudici}}, \bibinfo {author} {\bibfnamefont {H.}~\bibnamefont {Pichler}},\ and\ \bibinfo {author} {\bibfnamefont {A.~M.}\ \bibnamefont {Kaufman}},\ }\href {https://doi.org/10.1038/s41567-026-03258-8} {\bibfield  {journal} {\bibinfo  {journal} {Nat. Phys.}\ }\textbf {\bibinfo {volume} {22}},\ \bibinfo {pages} {903} (\bibinfo {year} {2026})}\BibitemShut {NoStop}%
\bibitem [{\citenamefont {Radnaev}\ \emph {et~al.}(2025)\citenamefont {Radnaev}, \citenamefont {Chung}, \citenamefont {Cole}, \citenamefont {Mason}, \citenamefont {Ballance}, \citenamefont {Bedalov}, \citenamefont {Belknap}, \citenamefont {Berman}, \citenamefont {Blakely}, \citenamefont {Bloomfield}, \citenamefont {Buttler}, \citenamefont {Campbell}, \citenamefont {Chopinaud}, \citenamefont {Copenhaver}, \citenamefont {Dawes}, \citenamefont {Eubanks}, \citenamefont {Friss}, \citenamefont {Garcia}, \citenamefont {Gilbert}, \citenamefont {Gillette}, \citenamefont {Goiporia}, \citenamefont {Gokhale}, \citenamefont {Goldwin}, \citenamefont {Goodwin}, \citenamefont {Graham}, \citenamefont {Guttormsson}, \citenamefont {Hickman}, \citenamefont {Hurtley}, \citenamefont {Iliev}, \citenamefont {Jones}, \citenamefont {Jones}, \citenamefont {Kuper}, \citenamefont {Lewis}, \citenamefont {Lichtman}, \citenamefont {Majdeteimouri}, \citenamefont {Mason}, \citenamefont {McMaster}, \citenamefont {Miles}, \citenamefont
  {Mitchell}, \citenamefont {Murphree}, \citenamefont {Neff-Mallon}, \citenamefont {Oh}, \citenamefont {Omole}, \citenamefont {Parlo~Simon}, \citenamefont {Pederson}, \citenamefont {Perlin}, \citenamefont {Reiter}, \citenamefont {Rines}, \citenamefont {Romlow}, \citenamefont {Scott}, \citenamefont {Stiefvater}, \citenamefont {Tanner}, \citenamefont {Tucker}, \citenamefont {Vinogradov}, \citenamefont {Warter}, \citenamefont {Yeo}, \citenamefont {Saffman},\ and\ \citenamefont {Noel}}]{99.3_Radnaev2025}%
  \BibitemOpen
  \bibfield  {author} {\bibinfo {author} {\bibfnamefont {A.}~\bibnamefont {Radnaev}}, \bibinfo {author} {\bibfnamefont {W.}~\bibnamefont {Chung}}, \bibinfo {author} {\bibfnamefont {D.}~\bibnamefont {Cole}}, \bibinfo {author} {\bibfnamefont {D.}~\bibnamefont {Mason}}, \bibinfo {author} {\bibfnamefont {T.}~\bibnamefont {Ballance}}, \bibinfo {author} {\bibfnamefont {M.}~\bibnamefont {Bedalov}}, \bibinfo {author} {\bibfnamefont {D.}~\bibnamefont {Belknap}}, \bibinfo {author} {\bibfnamefont {M.}~\bibnamefont {Berman}}, \bibinfo {author} {\bibfnamefont {M.}~\bibnamefont {Blakely}}, \bibinfo {author} {\bibfnamefont {I.}~\bibnamefont {Bloomfield}}, \bibinfo {author} {\bibfnamefont {P.}~\bibnamefont {Buttler}}, \bibinfo {author} {\bibfnamefont {C.}~\bibnamefont {Campbell}}, \bibinfo {author} {\bibfnamefont {A.}~\bibnamefont {Chopinaud}}, \bibinfo {author} {\bibfnamefont {E.}~\bibnamefont {Copenhaver}}, \bibinfo {author} {\bibfnamefont {M.}~\bibnamefont {Dawes}}, \bibinfo {author} {\bibfnamefont {S.}~\bibnamefont
  {Eubanks}}, \bibinfo {author} {\bibfnamefont {A.}~\bibnamefont {Friss}}, \bibinfo {author} {\bibfnamefont {D.}~\bibnamefont {Garcia}}, \bibinfo {author} {\bibfnamefont {J.}~\bibnamefont {Gilbert}}, \bibinfo {author} {\bibfnamefont {M.}~\bibnamefont {Gillette}}, \bibinfo {author} {\bibfnamefont {P.}~\bibnamefont {Goiporia}}, \bibinfo {author} {\bibfnamefont {P.}~\bibnamefont {Gokhale}}, \bibinfo {author} {\bibfnamefont {J.}~\bibnamefont {Goldwin}}, \bibinfo {author} {\bibfnamefont {D.}~\bibnamefont {Goodwin}}, \bibinfo {author} {\bibfnamefont {T.}~\bibnamefont {Graham}}, \bibinfo {author} {\bibfnamefont {C.}~\bibnamefont {Guttormsson}}, \bibinfo {author} {\bibfnamefont {G.}~\bibnamefont {Hickman}}, \bibinfo {author} {\bibfnamefont {L.}~\bibnamefont {Hurtley}}, \bibinfo {author} {\bibfnamefont {M.}~\bibnamefont {Iliev}}, \bibinfo {author} {\bibfnamefont {E.}~\bibnamefont {Jones}}, \bibinfo {author} {\bibfnamefont {R.}~\bibnamefont {Jones}}, \bibinfo {author} {\bibfnamefont {K.}~\bibnamefont {Kuper}}, \bibinfo
  {author} {\bibfnamefont {T.}~\bibnamefont {Lewis}}, \bibinfo {author} {\bibfnamefont {M.}~\bibnamefont {Lichtman}}, \bibinfo {author} {\bibfnamefont {F.}~\bibnamefont {Majdeteimouri}}, \bibinfo {author} {\bibfnamefont {J.}~\bibnamefont {Mason}}, \bibinfo {author} {\bibfnamefont {J.}~\bibnamefont {McMaster}}, \bibinfo {author} {\bibfnamefont {J.}~\bibnamefont {Miles}}, \bibinfo {author} {\bibfnamefont {P.}~\bibnamefont {Mitchell}}, \bibinfo {author} {\bibfnamefont {J.}~\bibnamefont {Murphree}}, \bibinfo {author} {\bibfnamefont {N.}~\bibnamefont {Neff-Mallon}}, \bibinfo {author} {\bibfnamefont {T.}~\bibnamefont {Oh}}, \bibinfo {author} {\bibfnamefont {V.}~\bibnamefont {Omole}}, \bibinfo {author} {\bibfnamefont {C.}~\bibnamefont {Parlo~Simon}}, \bibinfo {author} {\bibfnamefont {N.}~\bibnamefont {Pederson}}, \bibinfo {author} {\bibfnamefont {M.}~\bibnamefont {Perlin}}, \bibinfo {author} {\bibfnamefont {A.}~\bibnamefont {Reiter}}, \bibinfo {author} {\bibfnamefont {R.}~\bibnamefont {Rines}}, \bibinfo {author}
  {\bibfnamefont {P.}~\bibnamefont {Romlow}}, \bibinfo {author} {\bibfnamefont {A.}~\bibnamefont {Scott}}, \bibinfo {author} {\bibfnamefont {D.}~\bibnamefont {Stiefvater}}, \bibinfo {author} {\bibfnamefont {J.}~\bibnamefont {Tanner}}, \bibinfo {author} {\bibfnamefont {A.}~\bibnamefont {Tucker}}, \bibinfo {author} {\bibfnamefont {I.}~\bibnamefont {Vinogradov}}, \bibinfo {author} {\bibfnamefont {M.}~\bibnamefont {Warter}}, \bibinfo {author} {\bibfnamefont {M.}~\bibnamefont {Yeo}}, \bibinfo {author} {\bibfnamefont {M.}~\bibnamefont {Saffman}},\ and\ \bibinfo {author} {\bibfnamefont {T.}~\bibnamefont {Noel}},\ }\href {https://doi.org/10.1103/66s8-jj18} {\bibfield  {journal} {\bibinfo  {journal} {PRX Quantum}\ }\textbf {\bibinfo {volume} {6}},\ \bibinfo {pages} {030334} (\bibinfo {year} {2025})}\BibitemShut {NoStop}%
\bibitem [{\citenamefont {Muniz}\ \emph {et~al.}(2025{\natexlab{b}})\citenamefont {Muniz}, \citenamefont {Stone}, \citenamefont {Stack}, \citenamefont {Jaffe}, \citenamefont {Kindem}, \citenamefont {Wadleigh}, \citenamefont {Zalys-Geller}, \citenamefont {Zhang}, \citenamefont {Chen}, \citenamefont {Norcia}, \citenamefont {Epstein}, \citenamefont {Halperin}, \citenamefont {Hummel}, \citenamefont {Wilkason}, \citenamefont {Li}, \citenamefont {Barnes}, \citenamefont {Battaglino}, \citenamefont {Bohdanowicz}, \citenamefont {Booth}, \citenamefont {Brown}, \citenamefont {Brown}, \citenamefont {Cairncross}, \citenamefont {Cassella}, \citenamefont {Coxe}, \citenamefont {Crow}, \citenamefont {Feldkamp}, \citenamefont {Griger}, \citenamefont {Heinz}, \citenamefont {Jones}, \citenamefont {Kim}, \citenamefont {King}, \citenamefont {Kotru}, \citenamefont {Lauigan}, \citenamefont {Marjanovic}, \citenamefont {Megidish}, \citenamefont {Meredith}, \citenamefont {McDonald}, \citenamefont {Morshead}, \citenamefont
  {Narayanaswami}, \citenamefont {Nishiguchi}, \citenamefont {Paule}, \citenamefont {Pawlak}, \citenamefont {Pudenz}, \citenamefont {P\'erez}, \citenamefont {Ryou}, \citenamefont {Simon}, \citenamefont {Smull}, \citenamefont {Urbanek}, \citenamefont {van~de Veerdonk}, \citenamefont {Vendeiro}, \citenamefont {Wu}, \citenamefont {Xie},\ and\ \citenamefont {Bloom}}]{AtomComputing2025Gate}%
  \BibitemOpen
  \bibfield  {author} {\bibinfo {author} {\bibfnamefont {J.~A.}\ \bibnamefont {Muniz}}, \bibinfo {author} {\bibfnamefont {M.}~\bibnamefont {Stone}}, \bibinfo {author} {\bibfnamefont {D.~T.}\ \bibnamefont {Stack}}, \bibinfo {author} {\bibfnamefont {M.}~\bibnamefont {Jaffe}}, \bibinfo {author} {\bibfnamefont {J.~M.}\ \bibnamefont {Kindem}}, \bibinfo {author} {\bibfnamefont {L.}~\bibnamefont {Wadleigh}}, \bibinfo {author} {\bibfnamefont {E.}~\bibnamefont {Zalys-Geller}}, \bibinfo {author} {\bibfnamefont {X.}~\bibnamefont {Zhang}}, \bibinfo {author} {\bibfnamefont {C.-A.}\ \bibnamefont {Chen}}, \bibinfo {author} {\bibfnamefont {M.~A.}\ \bibnamefont {Norcia}}, \bibinfo {author} {\bibfnamefont {J.}~\bibnamefont {Epstein}}, \bibinfo {author} {\bibfnamefont {E.}~\bibnamefont {Halperin}}, \bibinfo {author} {\bibfnamefont {F.}~\bibnamefont {Hummel}}, \bibinfo {author} {\bibfnamefont {T.}~\bibnamefont {Wilkason}}, \bibinfo {author} {\bibfnamefont {M.}~\bibnamefont {Li}}, \bibinfo {author} {\bibfnamefont
  {K.}~\bibnamefont {Barnes}}, \bibinfo {author} {\bibfnamefont {P.}~\bibnamefont {Battaglino}}, \bibinfo {author} {\bibfnamefont {T.~C.}\ \bibnamefont {Bohdanowicz}}, \bibinfo {author} {\bibfnamefont {G.}~\bibnamefont {Booth}}, \bibinfo {author} {\bibfnamefont {A.}~\bibnamefont {Brown}}, \bibinfo {author} {\bibfnamefont {M.~O.}\ \bibnamefont {Brown}}, \bibinfo {author} {\bibfnamefont {W.~B.}\ \bibnamefont {Cairncross}}, \bibinfo {author} {\bibfnamefont {K.}~\bibnamefont {Cassella}}, \bibinfo {author} {\bibfnamefont {R.}~\bibnamefont {Coxe}}, \bibinfo {author} {\bibfnamefont {D.}~\bibnamefont {Crow}}, \bibinfo {author} {\bibfnamefont {M.}~\bibnamefont {Feldkamp}}, \bibinfo {author} {\bibfnamefont {C.}~\bibnamefont {Griger}}, \bibinfo {author} {\bibfnamefont {A.}~\bibnamefont {Heinz}}, \bibinfo {author} {\bibfnamefont {A.~M.~W.}\ \bibnamefont {Jones}}, \bibinfo {author} {\bibfnamefont {H.}~\bibnamefont {Kim}}, \bibinfo {author} {\bibfnamefont {J.}~\bibnamefont {King}}, \bibinfo {author} {\bibfnamefont
  {K.}~\bibnamefont {Kotru}}, \bibinfo {author} {\bibfnamefont {J.}~\bibnamefont {Lauigan}}, \bibinfo {author} {\bibfnamefont {J.}~\bibnamefont {Marjanovic}}, \bibinfo {author} {\bibfnamefont {E.}~\bibnamefont {Megidish}}, \bibinfo {author} {\bibfnamefont {M.}~\bibnamefont {Meredith}}, \bibinfo {author} {\bibfnamefont {M.}~\bibnamefont {McDonald}}, \bibinfo {author} {\bibfnamefont {R.}~\bibnamefont {Morshead}}, \bibinfo {author} {\bibfnamefont {S.}~\bibnamefont {Narayanaswami}}, \bibinfo {author} {\bibfnamefont {C.}~\bibnamefont {Nishiguchi}}, \bibinfo {author} {\bibfnamefont {T.}~\bibnamefont {Paule}}, \bibinfo {author} {\bibfnamefont {K.~A.}\ \bibnamefont {Pawlak}}, \bibinfo {author} {\bibfnamefont {K.~L.}\ \bibnamefont {Pudenz}}, \bibinfo {author} {\bibfnamefont {D.~R.}\ \bibnamefont {P\'erez}}, \bibinfo {author} {\bibfnamefont {A.}~\bibnamefont {Ryou}}, \bibinfo {author} {\bibfnamefont {J.}~\bibnamefont {Simon}}, \bibinfo {author} {\bibfnamefont {A.}~\bibnamefont {Smull}}, \bibinfo {author} {\bibfnamefont
  {M.}~\bibnamefont {Urbanek}}, \bibinfo {author} {\bibfnamefont {R.~J.~M.}\ \bibnamefont {van~de Veerdonk}}, \bibinfo {author} {\bibfnamefont {Z.}~\bibnamefont {Vendeiro}}, \bibinfo {author} {\bibfnamefont {T.-Y.}\ \bibnamefont {Wu}}, \bibinfo {author} {\bibfnamefont {X.}~\bibnamefont {Xie}},\ and\ \bibinfo {author} {\bibfnamefont {B.~J.}\ \bibnamefont {Bloom}},\ }\href {https://doi.org/10.1103/PRXQuantum.6.020334} {\bibfield  {journal} {\bibinfo  {journal} {PRX Quantum}\ }\textbf {\bibinfo {volume} {6}},\ \bibinfo {pages} {020334} (\bibinfo {year} {2025}{\natexlab{b}})}\BibitemShut {NoStop}%
\bibitem [{\citenamefont {Lin}\ \emph {et~al.}(2026)\citenamefont {Lin}, \citenamefont {Li}, \citenamefont {Zheng}, \citenamefont {Hu}, \citenamefont {Chen}, \citenamefont {Wu}, \citenamefont {Zhang}, \citenamefont {Cheng}, \citenamefont {Deng}, \citenamefont {Wu}, \citenamefont {Chen}, \citenamefont {Rui}, \citenamefont {Lu},\ and\ \citenamefont {Pan}}]{99.60_lin2026}%
  \BibitemOpen
  \bibfield  {author} {\bibinfo {author} {\bibfnamefont {R.}~\bibnamefont {Lin}}, \bibinfo {author} {\bibfnamefont {Y.}~\bibnamefont {Li}}, \bibinfo {author} {\bibfnamefont {L.-T.}\ \bibnamefont {Zheng}}, \bibinfo {author} {\bibfnamefont {T.-R.}\ \bibnamefont {Hu}}, \bibinfo {author} {\bibfnamefont {S.-Y.}\ \bibnamefont {Chen}}, \bibinfo {author} {\bibfnamefont {H.-M.}\ \bibnamefont {Wu}}, \bibinfo {author} {\bibfnamefont {Y.-C.}\ \bibnamefont {Zhang}}, \bibinfo {author} {\bibfnamefont {H.-W.}\ \bibnamefont {Cheng}}, \bibinfo {author} {\bibfnamefont {Y.-H.}\ \bibnamefont {Deng}}, \bibinfo {author} {\bibfnamefont {Z.}~\bibnamefont {Wu}}, \bibinfo {author} {\bibfnamefont {M.-C.}\ \bibnamefont {Chen}}, \bibinfo {author} {\bibfnamefont {J.}~\bibnamefont {Rui}}, \bibinfo {author} {\bibfnamefont {C.-Y.}\ \bibnamefont {Lu}},\ and\ \bibinfo {author} {\bibfnamefont {J.-W.}\ \bibnamefont {Pan}},\ }\href {https://arxiv.org/abs/2603.01612} {\bibfield  {journal} {\bibinfo  {journal} {arXiv:2603.01612}\ } (\bibinfo {year}
  {2026})}\BibitemShut {NoStop}%
\bibitem [{\citenamefont {Lib}\ \emph {et~al.}(2026)\citenamefont {Lib}, \citenamefont {Timme}, \citenamefont {Ammenwerth}, \citenamefont {Gyger}, \citenamefont {Tao}, \citenamefont {Sun}, \citenamefont {Bloch},\ and\ \citenamefont {Zeiher}}]{Lib2026}%
  \BibitemOpen
  \bibfield  {author} {\bibinfo {author} {\bibfnamefont {O.}~\bibnamefont {Lib}}, \bibinfo {author} {\bibfnamefont {H.}~\bibnamefont {Timme}}, \bibinfo {author} {\bibfnamefont {M.}~\bibnamefont {Ammenwerth}}, \bibinfo {author} {\bibfnamefont {F.}~\bibnamefont {Gyger}}, \bibinfo {author} {\bibfnamefont {R.}~\bibnamefont {Tao}}, \bibinfo {author} {\bibfnamefont {S.}~\bibnamefont {Sun}}, \bibinfo {author} {\bibfnamefont {I.}~\bibnamefont {Bloch}},\ and\ \bibinfo {author} {\bibfnamefont {J.}~\bibnamefont {Zeiher}},\ }\href {https://arxiv.org/abs/2603.15561} {\bibfield  {journal} {\bibinfo  {journal} {arXiv:2603.15561}\ } (\bibinfo {year} {2026})}\BibitemShut {NoStop}%
\bibitem [{\citenamefont {Evered}\ \emph {et~al.}(2026)\citenamefont {Evered}, \citenamefont {Xu}, \citenamefont {Li}, \citenamefont {Geim}, \citenamefont {Bonilla~Ataides}, \citenamefont {Kalinowski}, \citenamefont {Bluvstein}, \citenamefont {Maskara}, \citenamefont {Kokail}, \citenamefont {Greiner}, \citenamefont {Vuleti{\'c}},\ and\ \citenamefont {Lukin}}]{99.85_evered2026}%
  \BibitemOpen
  \bibfield  {author} {\bibinfo {author} {\bibfnamefont {S.~J.}\ \bibnamefont {Evered}}, \bibinfo {author} {\bibfnamefont {M.}~\bibnamefont {Xu}}, \bibinfo {author} {\bibfnamefont {S.~H.}\ \bibnamefont {Li}}, \bibinfo {author} {\bibfnamefont {A.~A.}\ \bibnamefont {Geim}}, \bibinfo {author} {\bibfnamefont {J.~P.}\ \bibnamefont {Bonilla~Ataides}}, \bibinfo {author} {\bibfnamefont {M.}~\bibnamefont {Kalinowski}}, \bibinfo {author} {\bibfnamefont {D.}~\bibnamefont {Bluvstein}}, \bibinfo {author} {\bibfnamefont {N.}~\bibnamefont {Maskara}}, \bibinfo {author} {\bibfnamefont {C.}~\bibnamefont {Kokail}}, \bibinfo {author} {\bibfnamefont {M.}~\bibnamefont {Greiner}}, \bibinfo {author} {\bibfnamefont {V.}~\bibnamefont {Vuleti{\'c}}},\ and\ \bibinfo {author} {\bibfnamefont {M.~D.}\ \bibnamefont {Lukin}},\ }\href {https://doi.org/10.1126/science.aei1312} {\bibfield  {journal} {\bibinfo  {journal} {Science}\ }\textbf {\bibinfo {volume} {394}},\ \bibinfo {pages} {197} (\bibinfo {year} {2026})}\BibitemShut {NoStop}%
\bibitem [{\citenamefont {Saffman}(2016)}]{Saffman2016}%
  \BibitemOpen
  \bibfield  {author} {\bibinfo {author} {\bibfnamefont {M.}~\bibnamefont {Saffman}},\ }\href {https://doi.org/10.1088/0953-4075/49/20/202001} {\bibfield  {journal} {\bibinfo  {journal} {J. Phys. B: At. Mol. Opt. Phys.}\ }\textbf {\bibinfo {volume} {49}},\ \bibinfo {pages} {202001} (\bibinfo {year} {2016})}\BibitemShut {NoStop}%
\bibitem [{\citenamefont {Beterov}\ \emph {et~al.}(2009)\citenamefont {Beterov}, \citenamefont {Ryabtsev}, \citenamefont {Tretyakov},\ and\ \citenamefont {Entin}}]{Beterov2009}%
  \BibitemOpen
  \bibfield  {author} {\bibinfo {author} {\bibfnamefont {I.~I.}\ \bibnamefont {Beterov}}, \bibinfo {author} {\bibfnamefont {I.~I.}\ \bibnamefont {Ryabtsev}}, \bibinfo {author} {\bibfnamefont {D.~B.}\ \bibnamefont {Tretyakov}},\ and\ \bibinfo {author} {\bibfnamefont {V.~M.}\ \bibnamefont {Entin}},\ }\href {https://doi.org/10.1103/PhysRevA.79.052504} {\bibfield  {journal} {\bibinfo  {journal} {Phys. Rev. A}\ }\textbf {\bibinfo {volume} {79}},\ \bibinfo {pages} {052504} (\bibinfo {year} {2009})}\BibitemShut {NoStop}%
\bibitem [{\citenamefont {Goldschmidt}\ \emph {et~al.}(2016)\citenamefont {Goldschmidt}, \citenamefont {Boulier}, \citenamefont {Brown}, \citenamefont {Koller}, \citenamefont {Young}, \citenamefont {Gorshkov}, \citenamefont {Rolston},\ and\ \citenamefont {Porto}}]{Broadening_Goldschmidt2016}%
  \BibitemOpen
  \bibfield  {author} {\bibinfo {author} {\bibfnamefont {E.~A.}\ \bibnamefont {Goldschmidt}}, \bibinfo {author} {\bibfnamefont {T.}~\bibnamefont {Boulier}}, \bibinfo {author} {\bibfnamefont {R.~C.}\ \bibnamefont {Brown}}, \bibinfo {author} {\bibfnamefont {S.~B.}\ \bibnamefont {Koller}}, \bibinfo {author} {\bibfnamefont {J.~T.}\ \bibnamefont {Young}}, \bibinfo {author} {\bibfnamefont {A.~V.}\ \bibnamefont {Gorshkov}}, \bibinfo {author} {\bibfnamefont {S.~L.}\ \bibnamefont {Rolston}},\ and\ \bibinfo {author} {\bibfnamefont {J.~V.}\ \bibnamefont {Porto}},\ }\href {https://doi.org/10.1103/PhysRevLett.116.113001} {\bibfield  {journal} {\bibinfo  {journal} {Phys. Rev. Lett.}\ }\textbf {\bibinfo {volume} {116}},\ \bibinfo {pages} {113001} (\bibinfo {year} {2016})}\BibitemShut {NoStop}%
\bibitem [{\citenamefont {Festa}\ \emph {et~al.}(2022)\citenamefont {Festa}, \citenamefont {Lorenz}, \citenamefont {Steinert}, \citenamefont {Chen}, \citenamefont {Osterholz}, \citenamefont {Eberhard},\ and\ \citenamefont {Gross}}]{Festa2022}%
  \BibitemOpen
  \bibfield  {author} {\bibinfo {author} {\bibfnamefont {L.}~\bibnamefont {Festa}}, \bibinfo {author} {\bibfnamefont {N.}~\bibnamefont {Lorenz}}, \bibinfo {author} {\bibfnamefont {L.-M.}\ \bibnamefont {Steinert}}, \bibinfo {author} {\bibfnamefont {Z.}~\bibnamefont {Chen}}, \bibinfo {author} {\bibfnamefont {P.}~\bibnamefont {Osterholz}}, \bibinfo {author} {\bibfnamefont {R.}~\bibnamefont {Eberhard}},\ and\ \bibinfo {author} {\bibfnamefont {C.}~\bibnamefont {Gross}},\ }\href {https://doi.org/10.1103/PhysRevA.105.013109} {\bibfield  {journal} {\bibinfo  {journal} {Phys. Rev. A}\ }\textbf {\bibinfo {volume} {105}},\ \bibinfo {pages} {013109} (\bibinfo {year} {2022})}\BibitemShut {NoStop}%
\bibitem [{\citenamefont {Gallagher}\ and\ \citenamefont {Cooke}(1979)}]{Gallagher1979BBRTheory}%
  \BibitemOpen
  \bibfield  {author} {\bibinfo {author} {\bibfnamefont {T.~F.}\ \bibnamefont {Gallagher}}\ and\ \bibinfo {author} {\bibfnamefont {W.~E.}\ \bibnamefont {Cooke}},\ }\href {https://doi.org/10.1103/PhysRevLett.42.835} {\bibfield  {journal} {\bibinfo  {journal} {Phys. Rev. Lett.}\ }\textbf {\bibinfo {volume} {42}},\ \bibinfo {pages} {835} (\bibinfo {year} {1979})}\BibitemShut {NoStop}%
\bibitem [{\citenamefont {Spencer}\ \emph {et~al.}(1981)\citenamefont {Spencer}, \citenamefont {Vaidyanathan}, \citenamefont {Kleppner},\ and\ \citenamefont {Ducas}}]{Spencer1981Na}%
  \BibitemOpen
  \bibfield  {author} {\bibinfo {author} {\bibfnamefont {W.~P.}\ \bibnamefont {Spencer}}, \bibinfo {author} {\bibfnamefont {A.~G.}\ \bibnamefont {Vaidyanathan}}, \bibinfo {author} {\bibfnamefont {D.}~\bibnamefont {Kleppner}},\ and\ \bibinfo {author} {\bibfnamefont {T.~W.}\ \bibnamefont {Ducas}},\ }\href {https://doi.org/10.1103/PhysRevA.24.2513} {\bibfield  {journal} {\bibinfo  {journal} {Phys. Rev. A}\ }\textbf {\bibinfo {volume} {24}},\ \bibinfo {pages} {2513} (\bibinfo {year} {1981})}\BibitemShut {NoStop}%
\bibitem [{\citenamefont {Spencer}\ \emph {et~al.}(1982)\citenamefont {Spencer}, \citenamefont {Vaidyanathan}, \citenamefont {Kleppner},\ and\ \citenamefont {Ducas}}]{Spencer1982Na}%
  \BibitemOpen
  \bibfield  {author} {\bibinfo {author} {\bibfnamefont {W.~P.}\ \bibnamefont {Spencer}}, \bibinfo {author} {\bibfnamefont {A.~G.}\ \bibnamefont {Vaidyanathan}}, \bibinfo {author} {\bibfnamefont {D.}~\bibnamefont {Kleppner}},\ and\ \bibinfo {author} {\bibfnamefont {T.~W.}\ \bibnamefont {Ducas}},\ }\href {https://doi.org/10.1103/PhysRevA.25.380} {\bibfield  {journal} {\bibinfo  {journal} {Phys. Rev. A}\ }\textbf {\bibinfo {volume} {25}},\ \bibinfo {pages} {380} (\bibinfo {year} {1982})}\BibitemShut {NoStop}%
\bibitem [{\citenamefont {Cantat-Moltrecht}\ \emph {et~al.}(2020)\citenamefont {Cantat-Moltrecht}, \citenamefont {Corti\~nas}, \citenamefont {Ravon}, \citenamefont {M\'ehaignerie}, \citenamefont {Haroche}, \citenamefont {Raimond}, \citenamefont {Favier}, \citenamefont {Brune},\ and\ \citenamefont {Sayrin}}]{cantat2020long}%
  \BibitemOpen
  \bibfield  {author} {\bibinfo {author} {\bibfnamefont {T.}~\bibnamefont {Cantat-Moltrecht}}, \bibinfo {author} {\bibfnamefont {R.}~\bibnamefont {Corti\~nas}}, \bibinfo {author} {\bibfnamefont {B.}~\bibnamefont {Ravon}}, \bibinfo {author} {\bibfnamefont {P.}~\bibnamefont {M\'ehaignerie}}, \bibinfo {author} {\bibfnamefont {S.}~\bibnamefont {Haroche}}, \bibinfo {author} {\bibfnamefont {J.~M.}\ \bibnamefont {Raimond}}, \bibinfo {author} {\bibfnamefont {M.}~\bibnamefont {Favier}}, \bibinfo {author} {\bibfnamefont {M.}~\bibnamefont {Brune}},\ and\ \bibinfo {author} {\bibfnamefont {C.}~\bibnamefont {Sayrin}},\ }\href {https://doi.org/10.1103/PhysRevResearch.2.022032} {\bibfield  {journal} {\bibinfo  {journal} {Phys. Rev. Res.}\ }\textbf {\bibinfo {volume} {2}},\ \bibinfo {pages} {022032} (\bibinfo {year} {2020})}\BibitemShut {NoStop}%
\bibitem [{\citenamefont {Hulet}\ \emph {et~al.}(1985)\citenamefont {Hulet}, \citenamefont {Hilfer},\ and\ \citenamefont {Kleppner}}]{Hulet1985}%
  \BibitemOpen
  \bibfield  {author} {\bibinfo {author} {\bibfnamefont {R.~G.}\ \bibnamefont {Hulet}}, \bibinfo {author} {\bibfnamefont {E.~S.}\ \bibnamefont {Hilfer}},\ and\ \bibinfo {author} {\bibfnamefont {D.}~\bibnamefont {Kleppner}},\ }\href {https://doi.org/10.1103/PhysRevLett.55.2137} {\bibfield  {journal} {\bibinfo  {journal} {Phys. Rev. Lett.}\ }\textbf {\bibinfo {volume} {55}},\ \bibinfo {pages} {2137} (\bibinfo {year} {1985})}\BibitemShut {NoStop}%
\bibitem [{\citenamefont {Nguyen}\ \emph {et~al.}(2018)\citenamefont {Nguyen}, \citenamefont {Raimond}, \citenamefont {Sayrin}, \citenamefont {Corti\~nas}, \citenamefont {Cantat-Moltrecht}, \citenamefont {Assemat}, \citenamefont {Dotsenko}, \citenamefont {Gleyzes}, \citenamefont {Haroche}, \citenamefont {Roux}, \citenamefont {Jolicoeur},\ and\ \citenamefont {Brune}}]{circularRydberg_Nguyen2018}%
  \BibitemOpen
  \bibfield  {author} {\bibinfo {author} {\bibfnamefont {T.~L.}\ \bibnamefont {Nguyen}}, \bibinfo {author} {\bibfnamefont {J.~M.}\ \bibnamefont {Raimond}}, \bibinfo {author} {\bibfnamefont {C.}~\bibnamefont {Sayrin}}, \bibinfo {author} {\bibfnamefont {R.}~\bibnamefont {Corti\~nas}}, \bibinfo {author} {\bibfnamefont {T.}~\bibnamefont {Cantat-Moltrecht}}, \bibinfo {author} {\bibfnamefont {F.}~\bibnamefont {Assemat}}, \bibinfo {author} {\bibfnamefont {I.}~\bibnamefont {Dotsenko}}, \bibinfo {author} {\bibfnamefont {S.}~\bibnamefont {Gleyzes}}, \bibinfo {author} {\bibfnamefont {S.}~\bibnamefont {Haroche}}, \bibinfo {author} {\bibfnamefont {G.}~\bibnamefont {Roux}}, \bibinfo {author} {\bibfnamefont {T.}~\bibnamefont {Jolicoeur}},\ and\ \bibinfo {author} {\bibfnamefont {M.}~\bibnamefont {Brune}},\ }\href {https://doi.org/10.1103/PhysRevX.8.011032} {\bibfield  {journal} {\bibinfo  {journal} {Phys. Rev. X}\ }\textbf {\bibinfo {volume} {8}},\ \bibinfo {pages} {011032} (\bibinfo {year} {2018})}\BibitemShut {NoStop}%
\bibitem [{\citenamefont {Cohen}\ and\ \citenamefont {Thompson}(2021)}]{circularRydberg_Cohen2021}%
  \BibitemOpen
  \bibfield  {author} {\bibinfo {author} {\bibfnamefont {S.~R.}\ \bibnamefont {Cohen}}\ and\ \bibinfo {author} {\bibfnamefont {J.~D.}\ \bibnamefont {Thompson}},\ }\href {https://doi.org/10.1103/PRXQuantum.2.030322} {\bibfield  {journal} {\bibinfo  {journal} {PRX Quantum}\ }\textbf {\bibinfo {volume} {2}},\ \bibinfo {pages} {030322} (\bibinfo {year} {2021})}\BibitemShut {NoStop}%
\bibitem [{\citenamefont {Wu}\ \emph {et~al.}(2023)\citenamefont {Wu}, \citenamefont {Richaud}, \citenamefont {Raimond}, \citenamefont {Brune},\ and\ \citenamefont {Gleyzes}}]{circular-Rydberg_Wu2023}%
  \BibitemOpen
  \bibfield  {author} {\bibinfo {author} {\bibfnamefont {H.}~\bibnamefont {Wu}}, \bibinfo {author} {\bibfnamefont {R.}~\bibnamefont {Richaud}}, \bibinfo {author} {\bibfnamefont {J.-M.}\ \bibnamefont {Raimond}}, \bibinfo {author} {\bibfnamefont {M.}~\bibnamefont {Brune}},\ and\ \bibinfo {author} {\bibfnamefont {S.}~\bibnamefont {Gleyzes}},\ }\href {https://doi.org/10.1103/PhysRevLett.130.023202} {\bibfield  {journal} {\bibinfo  {journal} {Phys. Rev. Lett.}\ }\textbf {\bibinfo {volume} {130}},\ \bibinfo {pages} {023202} (\bibinfo {year} {2023})}\BibitemShut {NoStop}%
\bibitem [{\citenamefont {H\"olzl}\ \emph {et~al.}(2024)\citenamefont {H\"olzl}, \citenamefont {G\"otzelmann}, \citenamefont {Pultinevicius}, \citenamefont {Wirth},\ and\ \citenamefont {Meinert}}]{Circular_Sr_Holzl2024}%
  \BibitemOpen
  \bibfield  {author} {\bibinfo {author} {\bibfnamefont {C.}~\bibnamefont {H\"olzl}}, \bibinfo {author} {\bibfnamefont {A.}~\bibnamefont {G\"otzelmann}}, \bibinfo {author} {\bibfnamefont {E.}~\bibnamefont {Pultinevicius}}, \bibinfo {author} {\bibfnamefont {M.}~\bibnamefont {Wirth}},\ and\ \bibinfo {author} {\bibfnamefont {F.}~\bibnamefont {Meinert}},\ }\href {https://doi.org/10.1103/PhysRevX.14.021024} {\bibfield  {journal} {\bibinfo  {journal} {Phys. Rev. X}\ }\textbf {\bibinfo {volume} {14}},\ \bibinfo {pages} {021024} (\bibinfo {year} {2024})}\BibitemShut {NoStop}%
\bibitem [{\citenamefont {Pultinevicius}\ \emph {et~al.}(2026)\citenamefont {Pultinevicius}, \citenamefont {G{\"o}tzelmann}, \citenamefont {Thielemann}, \citenamefont {H{\"o}lzl},\ and\ \citenamefont {Meinert}}]{pultinevicius2026long}%
  \BibitemOpen
  \bibfield  {author} {\bibinfo {author} {\bibfnamefont {E.}~\bibnamefont {Pultinevicius}}, \bibinfo {author} {\bibfnamefont {A.}~\bibnamefont {G{\"o}tzelmann}}, \bibinfo {author} {\bibfnamefont {F.}~\bibnamefont {Thielemann}}, \bibinfo {author} {\bibfnamefont {C.}~\bibnamefont {H{\"o}lzl}},\ and\ \bibinfo {author} {\bibfnamefont {F.}~\bibnamefont {Meinert}},\ }\href {https://www.nature.com/articles/s41467-026-77764-x} {\bibfield  {journal} {\bibinfo  {journal} {Nat. Commun.}\ }\textbf {\bibinfo {volume} {17}},\ \bibinfo {pages} {9834} (\bibinfo {year} {2026})}\BibitemShut {NoStop}%
\bibitem [{\citenamefont {Schymik}\ \emph {et~al.}(2021)\citenamefont {Schymik}, \citenamefont {Pancaldi}, \citenamefont {Nogrette}, \citenamefont {Barredo}, \citenamefont {Paris}, \citenamefont {Browaeys},\ and\ \citenamefont {Lahaye}}]{6000s_Schymik2021}%
  \BibitemOpen
  \bibfield  {author} {\bibinfo {author} {\bibfnamefont {K.-N.}\ \bibnamefont {Schymik}}, \bibinfo {author} {\bibfnamefont {S.}~\bibnamefont {Pancaldi}}, \bibinfo {author} {\bibfnamefont {F.}~\bibnamefont {Nogrette}}, \bibinfo {author} {\bibfnamefont {D.}~\bibnamefont {Barredo}}, \bibinfo {author} {\bibfnamefont {J.}~\bibnamefont {Paris}}, \bibinfo {author} {\bibfnamefont {A.}~\bibnamefont {Browaeys}},\ and\ \bibinfo {author} {\bibfnamefont {T.}~\bibnamefont {Lahaye}},\ }\href {https://doi.org/10.1103/PhysRevApplied.16.034013} {\bibfield  {journal} {\bibinfo  {journal} {Phys. Rev. Appl.}\ }\textbf {\bibinfo {volume} {16}},\ \bibinfo {pages} {034013} (\bibinfo {year} {2021})}\BibitemShut {NoStop}%
\bibitem [{\citenamefont {Zhang}\ \emph {et~al.}(2025)\citenamefont {Zhang}, \citenamefont {Hsu}, \citenamefont {Tan}, \citenamefont {Slichter}, \citenamefont {Kaufman}, \citenamefont {Marinelli},\ and\ \citenamefont {Regal}}]{3000s_Zhang2025}%
  \BibitemOpen
  \bibfield  {author} {\bibinfo {author} {\bibfnamefont {Z.}~\bibnamefont {Zhang}}, \bibinfo {author} {\bibfnamefont {T.-W.}\ \bibnamefont {Hsu}}, \bibinfo {author} {\bibfnamefont {T.~Y.}\ \bibnamefont {Tan}}, \bibinfo {author} {\bibfnamefont {D.~H.}\ \bibnamefont {Slichter}}, \bibinfo {author} {\bibfnamefont {A.~M.}\ \bibnamefont {Kaufman}}, \bibinfo {author} {\bibfnamefont {M.}~\bibnamefont {Marinelli}},\ and\ \bibinfo {author} {\bibfnamefont {C.~A.}\ \bibnamefont {Regal}},\ }\href {https://doi.org/10.1103/PRXQuantum.6.020337} {\bibfield  {journal} {\bibinfo  {journal} {PRX Quantum}\ }\textbf {\bibinfo {volume} {6}},\ \bibinfo {pages} {020337} (\bibinfo {year} {2025})}\BibitemShut {NoStop}%
\bibitem [{\citenamefont {Lim}\ \emph {et~al.}(2026)\citenamefont {Lim}, \citenamefont {Mamann}, \citenamefont {Pichard}, \citenamefont {Bourachot}, \citenamefont {Lindberg}, \citenamefont {Hamot}, \citenamefont {Le~Bars}, \citenamefont {Fasola}, \citenamefont {Tan}, \citenamefont {Cournez}, \citenamefont {Dutartre}, \citenamefont {Cartry}, \citenamefont {Lemettre}, \citenamefont {Hostein}, \citenamefont {Paris}, \citenamefont {Ferreyrol}, \citenamefont {Collardey}, \citenamefont {Signoles}, \citenamefont {Lahaye}, \citenamefont {Monmeyran},\ and\ \citenamefont {Ximenez}}]{defect-free_Lim2026}%
  \BibitemOpen
  \bibfield  {author} {\bibinfo {author} {\bibfnamefont {D.}~\bibnamefont {Lim}}, \bibinfo {author} {\bibfnamefont {H.}~\bibnamefont {Mamann}}, \bibinfo {author} {\bibfnamefont {G.}~\bibnamefont {Pichard}}, \bibinfo {author} {\bibfnamefont {L.}~\bibnamefont {Bourachot}}, \bibinfo {author} {\bibfnamefont {A.}~\bibnamefont {Lindberg}}, \bibinfo {author} {\bibfnamefont {C.}~\bibnamefont {Hamot}}, \bibinfo {author} {\bibfnamefont {H.}~\bibnamefont {Le~Bars}}, \bibinfo {author} {\bibfnamefont {F.}~\bibnamefont {Fasola}}, \bibinfo {author} {\bibfnamefont {S.}~\bibnamefont {Tan}}, \bibinfo {author} {\bibfnamefont {G.}~\bibnamefont {Cournez}}, \bibinfo {author} {\bibfnamefont {S.}~\bibnamefont {Dutartre}}, \bibinfo {author} {\bibfnamefont {T.}~\bibnamefont {Cartry}}, \bibinfo {author} {\bibfnamefont {S.}~\bibnamefont {Lemettre}}, \bibinfo {author} {\bibfnamefont {R.}~\bibnamefont {Hostein}}, \bibinfo {author} {\bibfnamefont {J.}~\bibnamefont {Paris}}, \bibinfo {author} {\bibfnamefont {F.}~\bibnamefont {Ferreyrol}},
  \bibinfo {author} {\bibfnamefont {A.}~\bibnamefont {Collardey}}, \bibinfo {author} {\bibfnamefont {A.}~\bibnamefont {Signoles}}, \bibinfo {author} {\bibfnamefont {T.}~\bibnamefont {Lahaye}}, \bibinfo {author} {\bibfnamefont {C.}~\bibnamefont {Monmeyran}},\ and\ \bibinfo {author} {\bibfnamefont {B.}~\bibnamefont {Ximenez}},\ }\href {https://doi.org/10.1103/45t1-vl5y} {\bibfield  {journal} {\bibinfo  {journal} {PRX Quantum}\ }\textbf {\bibinfo {volume} {7}},\ \bibinfo {pages} {033032} (\bibinfo {year} {2026})}\BibitemShut {NoStop}%
\bibitem [{\citenamefont {Jin}\ \emph {et~al.}(2026)\citenamefont {Jin}, \citenamefont {Shi}, \citenamefont {Alaoui}, \citenamefont {Deng}, \citenamefont {Lu}, \citenamefont {Thompson},\ and\ \citenamefont {Bakr}}]{lifetime_Jin2026}%
  \BibitemOpen
  \bibfield  {author} {\bibinfo {author} {\bibfnamefont {J.}~\bibnamefont {Jin}}, \bibinfo {author} {\bibfnamefont {Y.}~\bibnamefont {Shi}}, \bibinfo {author} {\bibfnamefont {Y.~A.}\ \bibnamefont {Alaoui}}, \bibinfo {author} {\bibfnamefont {J.}~\bibnamefont {Deng}}, \bibinfo {author} {\bibfnamefont {Y.}~\bibnamefont {Lu}}, \bibinfo {author} {\bibfnamefont {J.~D.}\ \bibnamefont {Thompson}},\ and\ \bibinfo {author} {\bibfnamefont {W.~S.}\ \bibnamefont {Bakr}},\ }\href {https://doi.org/10.1103/96bx-rjwz} {\bibfield  {journal} {\bibinfo  {journal} {PRX Quantum}\ }\textbf {\bibinfo {volume} {7}},\ \bibinfo {pages} {033009} (\bibinfo {year} {2026})}\BibitemShut {NoStop}%
\bibitem [{\citenamefont {Jau}\ \emph {et~al.}(2016)\citenamefont {Jau}, \citenamefont {Hankin}, \citenamefont {Keating}, \citenamefont {Deutsch},\ and\ \citenamefont {Biedermann}}]{dressing_Jau2016}%
  \BibitemOpen
  \bibfield  {author} {\bibinfo {author} {\bibfnamefont {Y.-Y.}\ \bibnamefont {Jau}}, \bibinfo {author} {\bibfnamefont {A.~M.}\ \bibnamefont {Hankin}}, \bibinfo {author} {\bibfnamefont {T.}~\bibnamefont {Keating}}, \bibinfo {author} {\bibfnamefont {I.~H.}\ \bibnamefont {Deutsch}},\ and\ \bibinfo {author} {\bibfnamefont {G.~W.}\ \bibnamefont {Biedermann}},\ }\href {https://doi.org/10.1038/nphys3487} {\bibfield  {journal} {\bibinfo  {journal} {Nat. Phys.}\ }\textbf {\bibinfo {volume} {12}},\ \bibinfo {pages} {71} (\bibinfo {year} {2016})}\BibitemShut {NoStop}%
\bibitem [{\citenamefont {Zeiher}\ \emph {et~al.}(2016)\citenamefont {Zeiher}, \citenamefont {van Bijnen}, \citenamefont {Schauß}, \citenamefont {Hild}, \citenamefont {Choi}, \citenamefont {Pohl}, \citenamefont {Bloch},\ and\ \citenamefont {Gross}}]{Dressing_Zeiher2016}%
  \BibitemOpen
  \bibfield  {author} {\bibinfo {author} {\bibfnamefont {J.}~\bibnamefont {Zeiher}}, \bibinfo {author} {\bibfnamefont {R.}~\bibnamefont {van Bijnen}}, \bibinfo {author} {\bibfnamefont {P.}~\bibnamefont {Schauß}}, \bibinfo {author} {\bibfnamefont {S.}~\bibnamefont {Hild}}, \bibinfo {author} {\bibfnamefont {J.-y.}\ \bibnamefont {Choi}}, \bibinfo {author} {\bibfnamefont {T.}~\bibnamefont {Pohl}}, \bibinfo {author} {\bibfnamefont {I.}~\bibnamefont {Bloch}},\ and\ \bibinfo {author} {\bibfnamefont {C.}~\bibnamefont {Gross}},\ }\href {https://doi.org/10.1038/nphys3835} {\bibfield  {journal} {\bibinfo  {journal} {Nat. Phys.}\ }\textbf {\bibinfo {volume} {12}},\ \bibinfo {pages} {1095} (\bibinfo {year} {2016})}\BibitemShut {NoStop}%
\bibitem [{\citenamefont {Hines}\ \emph {et~al.}(2023)\citenamefont {Hines}, \citenamefont {Rajagopal}, \citenamefont {Moreau}, \citenamefont {Wahrman}, \citenamefont {Lewis}, \citenamefont {Markovi\ifmmode~\acute{c}\else \'{c}\fi{}},\ and\ \citenamefont {Schleier-Smith}}]{Schleier-Smith2023DressingSqueezing}%
  \BibitemOpen
  \bibfield  {author} {\bibinfo {author} {\bibfnamefont {J.~A.}\ \bibnamefont {Hines}}, \bibinfo {author} {\bibfnamefont {S.~V.}\ \bibnamefont {Rajagopal}}, \bibinfo {author} {\bibfnamefont {G.~L.}\ \bibnamefont {Moreau}}, \bibinfo {author} {\bibfnamefont {M.~D.}\ \bibnamefont {Wahrman}}, \bibinfo {author} {\bibfnamefont {N.~A.}\ \bibnamefont {Lewis}}, \bibinfo {author} {\bibfnamefont {O.}~\bibnamefont {Markovi\ifmmode~\acute{c}\else \'{c}\fi{}}},\ and\ \bibinfo {author} {\bibfnamefont {M.}~\bibnamefont {Schleier-Smith}},\ }\href {https://doi.org/10.1103/PhysRevLett.131.063401} {\bibfield  {journal} {\bibinfo  {journal} {Phys. Rev. Lett.}\ }\textbf {\bibinfo {volume} {131}},\ \bibinfo {pages} {063401} (\bibinfo {year} {2023})}\BibitemShut {NoStop}%
\bibitem [{\citenamefont {Avila}\ \emph {et~al.}(1987)\citenamefont {Avila}, \citenamefont {Giordano}, \citenamefont {Candelier}, \citenamefont {de~Clercq}, \citenamefont {Theobald},\ and\ \citenamefont {Cerez}}]{OP_Avila1987}%
  \BibitemOpen
  \bibfield  {author} {\bibinfo {author} {\bibfnamefont {G.}~\bibnamefont {Avila}}, \bibinfo {author} {\bibfnamefont {V.}~\bibnamefont {Giordano}}, \bibinfo {author} {\bibfnamefont {V.}~\bibnamefont {Candelier}}, \bibinfo {author} {\bibfnamefont {E.}~\bibnamefont {de~Clercq}}, \bibinfo {author} {\bibfnamefont {G.}~\bibnamefont {Theobald}},\ and\ \bibinfo {author} {\bibfnamefont {P.}~\bibnamefont {Cerez}},\ }\href {https://doi.org/10.1103/PhysRevA.36.3719} {\bibfield  {journal} {\bibinfo  {journal} {Phys. Rev. A}\ }\textbf {\bibinfo {volume} {36}},\ \bibinfo {pages} {3719} (\bibinfo {year} {1987})}\BibitemShut {NoStop}%
\bibitem [{\citenamefont {McGuirk}\ \emph {et~al.}(2002)\citenamefont {McGuirk}, \citenamefont {Foster}, \citenamefont {Fixler}, \citenamefont {Snadden},\ and\ \citenamefont {Kasevich}}]{op_McGuirk2002}%
  \BibitemOpen
  \bibfield  {author} {\bibinfo {author} {\bibfnamefont {J.~M.}\ \bibnamefont {McGuirk}}, \bibinfo {author} {\bibfnamefont {G.~T.}\ \bibnamefont {Foster}}, \bibinfo {author} {\bibfnamefont {J.~B.}\ \bibnamefont {Fixler}}, \bibinfo {author} {\bibfnamefont {M.~J.}\ \bibnamefont {Snadden}},\ and\ \bibinfo {author} {\bibfnamefont {M.~A.}\ \bibnamefont {Kasevich}},\ }\href {https://doi.org/10.1103/PhysRevA.65.033608} {\bibfield  {journal} {\bibinfo  {journal} {Phys. Rev. A}\ }\textbf {\bibinfo {volume} {65}},\ \bibinfo {pages} {033608} (\bibinfo {year} {2002})}\BibitemShut {NoStop}%
\bibitem [{\citenamefont {Levine}\ \emph {et~al.}(2022)\citenamefont {Levine}, \citenamefont {Bluvstein}, \citenamefont {Keesling}, \citenamefont {Wang}, \citenamefont {Ebadi}, \citenamefont {Semeghini}, \citenamefont {Omran}, \citenamefont {Greiner}, \citenamefont {Vuleti{\ifmmode \acute{c}\else \'{c}\fi{}}},\ and\ \citenamefont {Lukin}}]{Raman_Levine2022}%
  \BibitemOpen
  \bibfield  {author} {\bibinfo {author} {\bibfnamefont {H.}~\bibnamefont {Levine}}, \bibinfo {author} {\bibfnamefont {D.}~\bibnamefont {Bluvstein}}, \bibinfo {author} {\bibfnamefont {A.}~\bibnamefont {Keesling}}, \bibinfo {author} {\bibfnamefont {T.~T.}\ \bibnamefont {Wang}}, \bibinfo {author} {\bibfnamefont {S.}~\bibnamefont {Ebadi}}, \bibinfo {author} {\bibfnamefont {G.}~\bibnamefont {Semeghini}}, \bibinfo {author} {\bibfnamefont {A.}~\bibnamefont {Omran}}, \bibinfo {author} {\bibfnamefont {M.}~\bibnamefont {Greiner}}, \bibinfo {author} {\bibfnamefont {V.}~\bibnamefont {Vuleti{\ifmmode \acute{c}\else \'{c}\fi{}}}},\ and\ \bibinfo {author} {\bibfnamefont {M.~D.}\ \bibnamefont {Lukin}},\ }\href {https://doi.org/10.1103/PhysRevA.105.032618} {\bibfield  {journal} {\bibinfo  {journal} {Phys. Rev. A}\ }\textbf {\bibinfo {volume} {105}},\ \bibinfo {pages} {032618} (\bibinfo {year} {2022})}\BibitemShut {NoStop}%
\bibitem [{\citenamefont {Kuhr}\ \emph {et~al.}(2005)\citenamefont {Kuhr}, \citenamefont {Alt}, \citenamefont {Schrader}, \citenamefont {Dotsenko}, \citenamefont {Miroshnychenko}, \citenamefont {Rauschenbeutel},\ and\ \citenamefont {Meschede}}]{dephasing_Kuhr2005}%
  \BibitemOpen
  \bibfield  {author} {\bibinfo {author} {\bibfnamefont {S.}~\bibnamefont {Kuhr}}, \bibinfo {author} {\bibfnamefont {W.}~\bibnamefont {Alt}}, \bibinfo {author} {\bibfnamefont {D.}~\bibnamefont {Schrader}}, \bibinfo {author} {\bibfnamefont {I.}~\bibnamefont {Dotsenko}}, \bibinfo {author} {\bibfnamefont {Y.}~\bibnamefont {Miroshnychenko}}, \bibinfo {author} {\bibfnamefont {A.}~\bibnamefont {Rauschenbeutel}},\ and\ \bibinfo {author} {\bibfnamefont {D.}~\bibnamefont {Meschede}},\ }\href {https://doi.org/10.1103/PhysRevA.72.023406} {\bibfield  {journal} {\bibinfo  {journal} {Phys. Rev. A}\ }\textbf {\bibinfo {volume} {72}},\ \bibinfo {pages} {023406} (\bibinfo {year} {2005})}\BibitemShut {NoStop}%
\bibitem [{\citenamefont {Thompson}\ \emph {et~al.}(2013)\citenamefont {Thompson}, \citenamefont {Tiecke}, \citenamefont {Zibrov}, \citenamefont {Vuleti{\ifmmode \acute{c}\else \'{c}\fi{}}},\ and\ \citenamefont {Lukin}}]{RSC_Thompson2013}%
  \BibitemOpen
  \bibfield  {author} {\bibinfo {author} {\bibfnamefont {J.~D.}\ \bibnamefont {Thompson}}, \bibinfo {author} {\bibfnamefont {T.~G.}\ \bibnamefont {Tiecke}}, \bibinfo {author} {\bibfnamefont {A.~S.}\ \bibnamefont {Zibrov}}, \bibinfo {author} {\bibfnamefont {V.}~\bibnamefont {Vuleti{\ifmmode \acute{c}\else \'{c}\fi{}}}},\ and\ \bibinfo {author} {\bibfnamefont {M.~D.}\ \bibnamefont {Lukin}},\ }\href {https://doi.org/10.1103/PhysRevLett.110.133001} {\bibfield  {journal} {\bibinfo  {journal} {Phys. Rev. Lett.}\ }\textbf {\bibinfo {volume} {110}},\ \bibinfo {pages} {133001} (\bibinfo {year} {2013})}\BibitemShut {NoStop}%
\bibitem [{\citenamefont {Gullion}\ \emph {et~al.}(1990)\citenamefont {Gullion}, \citenamefont {Baker},\ and\ \citenamefont {Conradi}}]{XY16_Gullion1990}%
  \BibitemOpen
  \bibfield  {author} {\bibinfo {author} {\bibfnamefont {T.}~\bibnamefont {Gullion}}, \bibinfo {author} {\bibfnamefont {D.~B.}\ \bibnamefont {Baker}},\ and\ \bibinfo {author} {\bibfnamefont {M.~S.}\ \bibnamefont {Conradi}},\ }\href {https://doi.org/10.1016/0022-2364(90)90331-3} {\bibfield  {journal} {\bibinfo  {journal} {J. Magn. Reson.}\ }\textbf {\bibinfo {volume} {89}},\ \bibinfo {pages} {479} (\bibinfo {year} {1990})}\BibitemShut {NoStop}%
\bibitem [{\citenamefont {Wimperis}(1994)}]{BB1_Wimperis1994}%
  \BibitemOpen
  \bibfield  {author} {\bibinfo {author} {\bibfnamefont {S.}~\bibnamefont {Wimperis}},\ }\href {https://doi.org/10.1006/jmra.1994.1159} {\bibfield  {journal} {\bibinfo  {journal} {J. Magn. Reson., Ser. A}\ }\textbf {\bibinfo {volume} {109}},\ \bibinfo {pages} {221} (\bibinfo {year} {1994})}\BibitemShut {NoStop}%
\bibitem [{\citenamefont {Vandersypen}\ and\ \citenamefont {Chuang}(2005)}]{NMR_Vandersypen2005}%
  \BibitemOpen
  \bibfield  {author} {\bibinfo {author} {\bibfnamefont {L.~M.~K.}\ \bibnamefont {Vandersypen}}\ and\ \bibinfo {author} {\bibfnamefont {I.~L.}\ \bibnamefont {Chuang}},\ }\href {https://doi.org/10.1103/RevModPhys.76.1037} {\bibfield  {journal} {\bibinfo  {journal} {Rev. Mod. Phys.}\ }\textbf {\bibinfo {volume} {76}},\ \bibinfo {pages} {1037} (\bibinfo {year} {2005})}\BibitemShut {NoStop}%
\bibitem [{\citenamefont {McKay}\ \emph {et~al.}(2017)\citenamefont {McKay}, \citenamefont {Wood}, \citenamefont {Sheldon}, \citenamefont {Chow},\ and\ \citenamefont {Gambetta}}]{virtualZ_McKay2017}%
  \BibitemOpen
  \bibfield  {author} {\bibinfo {author} {\bibfnamefont {D.~C.}\ \bibnamefont {McKay}}, \bibinfo {author} {\bibfnamefont {C.~J.}\ \bibnamefont {Wood}}, \bibinfo {author} {\bibfnamefont {S.}~\bibnamefont {Sheldon}}, \bibinfo {author} {\bibfnamefont {J.~M.}\ \bibnamefont {Chow}},\ and\ \bibinfo {author} {\bibfnamefont {J.~M.}\ \bibnamefont {Gambetta}},\ }\href {https://doi.org/10.1103/PhysRevA.96.022330} {\bibfield  {journal} {\bibinfo  {journal} {Phys. Rev. A}\ }\textbf {\bibinfo {volume} {96}},\ \bibinfo {pages} {022330} (\bibinfo {year} {2017})}\BibitemShut {NoStop}%
\bibitem [{\citenamefont {Knill}\ \emph {et~al.}(2008)\citenamefont {Knill}, \citenamefont {Leibfried}, \citenamefont {Reichle}, \citenamefont {Britton}, \citenamefont {Blakestad}, \citenamefont {Jost}, \citenamefont {Langer}, \citenamefont {Ozeri}, \citenamefont {Seidelin},\ and\ \citenamefont {Wineland}}]{1qrb_Knill2008}%
  \BibitemOpen
  \bibfield  {author} {\bibinfo {author} {\bibfnamefont {E.}~\bibnamefont {Knill}}, \bibinfo {author} {\bibfnamefont {D.}~\bibnamefont {Leibfried}}, \bibinfo {author} {\bibfnamefont {R.}~\bibnamefont {Reichle}}, \bibinfo {author} {\bibfnamefont {J.}~\bibnamefont {Britton}}, \bibinfo {author} {\bibfnamefont {R.~B.}\ \bibnamefont {Blakestad}}, \bibinfo {author} {\bibfnamefont {J.~D.}\ \bibnamefont {Jost}}, \bibinfo {author} {\bibfnamefont {C.}~\bibnamefont {Langer}}, \bibinfo {author} {\bibfnamefont {R.}~\bibnamefont {Ozeri}}, \bibinfo {author} {\bibfnamefont {S.}~\bibnamefont {Seidelin}},\ and\ \bibinfo {author} {\bibfnamefont {D.~J.}\ \bibnamefont {Wineland}},\ }\href {https://doi.org/10.1103/PhysRevA.77.012307} {\bibfield  {journal} {\bibinfo  {journal} {Phys. Rev. A}\ }\textbf {\bibinfo {volume} {77}},\ \bibinfo {pages} {012307} (\bibinfo {year} {2008})}\BibitemShut {NoStop}%
\bibitem [{\citenamefont {Magesan}\ \emph {et~al.}(2011)\citenamefont {Magesan}, \citenamefont {Gambetta},\ and\ \citenamefont {Emerson}}]{1qrb_Magesan2011}%
  \BibitemOpen
  \bibfield  {author} {\bibinfo {author} {\bibfnamefont {E.}~\bibnamefont {Magesan}}, \bibinfo {author} {\bibfnamefont {J.~M.}\ \bibnamefont {Gambetta}},\ and\ \bibinfo {author} {\bibfnamefont {J.}~\bibnamefont {Emerson}},\ }\href {https://doi.org/10.1103/PhysRevLett.106.180504} {\bibfield  {journal} {\bibinfo  {journal} {Phys. Rev. Lett.}\ }\textbf {\bibinfo {volume} {106}},\ \bibinfo {pages} {180504} (\bibinfo {year} {2011})}\BibitemShut {NoStop}%
\bibitem [{\citenamefont {Šibalić}\ \emph {et~al.}(2017)\citenamefont {Šibalić}, \citenamefont {Pritchard}, \citenamefont {Adams},\ and\ \citenamefont {Weatherill}}]{arc_sibalic2017}%
  \BibitemOpen
  \bibfield  {author} {\bibinfo {author} {\bibfnamefont {N.}~\bibnamefont {Šibalić}}, \bibinfo {author} {\bibfnamefont {J.~D.}\ \bibnamefont {Pritchard}}, \bibinfo {author} {\bibfnamefont {C.~S.}\ \bibnamefont {Adams}},\ and\ \bibinfo {author} {\bibfnamefont {K.~J.}\ \bibnamefont {Weatherill}},\ }\href {https://doi.org/10.1016/j.cpc.2017.06.015} {\bibfield  {journal} {\bibinfo  {journal} {Comput. Phys. Commun.}\ }\textbf {\bibinfo {volume} {220}},\ \bibinfo {pages} {319} (\bibinfo {year} {2017})}\BibitemShut {NoStop}%
\bibitem [{\citenamefont {Ga{\"e}tan}\ \emph {et~al.}(2009)\citenamefont {Ga{\"e}tan}, \citenamefont {Miroshnychenko}, \citenamefont {Wilk}, \citenamefont {Chotia}, \citenamefont {Viteau}, \citenamefont {Comparat}, \citenamefont {Pillet}, \citenamefont {Browaeys},\ and\ \citenamefont {Grangier}}]{blockade_Gaetan2009}%
  \BibitemOpen
  \bibfield  {author} {\bibinfo {author} {\bibfnamefont {A.}~\bibnamefont {Ga{\"e}tan}}, \bibinfo {author} {\bibfnamefont {Y.}~\bibnamefont {Miroshnychenko}}, \bibinfo {author} {\bibfnamefont {T.}~\bibnamefont {Wilk}}, \bibinfo {author} {\bibfnamefont {A.}~\bibnamefont {Chotia}}, \bibinfo {author} {\bibfnamefont {M.}~\bibnamefont {Viteau}}, \bibinfo {author} {\bibfnamefont {D.}~\bibnamefont {Comparat}}, \bibinfo {author} {\bibfnamefont {P.}~\bibnamefont {Pillet}}, \bibinfo {author} {\bibfnamefont {A.}~\bibnamefont {Browaeys}},\ and\ \bibinfo {author} {\bibfnamefont {P.}~\bibnamefont {Grangier}},\ }\href {https://doi.org/10.1038/nphys1183} {\bibfield  {journal} {\bibinfo  {journal} {Nat. Phys.}\ }\textbf {\bibinfo {volume} {5}},\ \bibinfo {pages} {115} (\bibinfo {year} {2009})}\BibitemShut {NoStop}%
\bibitem [{\citenamefont {Urban}\ \emph {et~al.}(2009)\citenamefont {Urban}, \citenamefont {Johnson}, \citenamefont {Henage}, \citenamefont {Isenhower}, \citenamefont {Yavuz}, \citenamefont {Walker},\ and\ \citenamefont {Saffman}}]{blockade_Urban2009}%
  \BibitemOpen
  \bibfield  {author} {\bibinfo {author} {\bibfnamefont {E.}~\bibnamefont {Urban}}, \bibinfo {author} {\bibfnamefont {T.~A.}\ \bibnamefont {Johnson}}, \bibinfo {author} {\bibfnamefont {T.}~\bibnamefont {Henage}}, \bibinfo {author} {\bibfnamefont {L.}~\bibnamefont {Isenhower}}, \bibinfo {author} {\bibfnamefont {D.~D.}\ \bibnamefont {Yavuz}}, \bibinfo {author} {\bibfnamefont {T.~G.}\ \bibnamefont {Walker}},\ and\ \bibinfo {author} {\bibfnamefont {M.}~\bibnamefont {Saffman}},\ }\href {https://doi.org/10.1038/nphys1178} {\bibfield  {journal} {\bibinfo  {journal} {Nat. Phys.}\ }\textbf {\bibinfo {volume} {5}},\ \bibinfo {pages} {110} (\bibinfo {year} {2009})}\BibitemShut {NoStop}%
\bibitem [{\citenamefont {Isenhower}\ \emph {et~al.}(2010)\citenamefont {Isenhower}, \citenamefont {Urban}, \citenamefont {Zhang}, \citenamefont {Gill}, \citenamefont {Henage}, \citenamefont {Johnson}, \citenamefont {Walker},\ and\ \citenamefont {Saffman}}]{CNOT_Isenhower2010}%
  \BibitemOpen
  \bibfield  {author} {\bibinfo {author} {\bibfnamefont {L.}~\bibnamefont {Isenhower}}, \bibinfo {author} {\bibfnamefont {E.}~\bibnamefont {Urban}}, \bibinfo {author} {\bibfnamefont {X.~L.}\ \bibnamefont {Zhang}}, \bibinfo {author} {\bibfnamefont {A.~T.}\ \bibnamefont {Gill}}, \bibinfo {author} {\bibfnamefont {T.}~\bibnamefont {Henage}}, \bibinfo {author} {\bibfnamefont {T.~A.}\ \bibnamefont {Johnson}}, \bibinfo {author} {\bibfnamefont {T.~G.}\ \bibnamefont {Walker}},\ and\ \bibinfo {author} {\bibfnamefont {M.}~\bibnamefont {Saffman}},\ }\href {https://doi.org/10.1103/PhysRevLett.104.010503} {\bibfield  {journal} {\bibinfo  {journal} {Phys. Rev. Lett.}\ }\textbf {\bibinfo {volume} {104}},\ \bibinfo {pages} {010503} (\bibinfo {year} {2010})}\BibitemShut {NoStop}%
\bibitem [{\citenamefont {Wilk}\ \emph {et~al.}(2010)\citenamefont {Wilk}, \citenamefont {Ga{\"e}tan}, \citenamefont {Evellin}, \citenamefont {Wolters}, \citenamefont {Miroshnychenko}, \citenamefont {Grangier},\ and\ \citenamefont {Browaeys}}]{entanglement_Wilk2010}%
  \BibitemOpen
  \bibfield  {author} {\bibinfo {author} {\bibfnamefont {T.}~\bibnamefont {Wilk}}, \bibinfo {author} {\bibfnamefont {A.}~\bibnamefont {Ga{\"e}tan}}, \bibinfo {author} {\bibfnamefont {C.}~\bibnamefont {Evellin}}, \bibinfo {author} {\bibfnamefont {J.}~\bibnamefont {Wolters}}, \bibinfo {author} {\bibfnamefont {Y.}~\bibnamefont {Miroshnychenko}}, \bibinfo {author} {\bibfnamefont {P.}~\bibnamefont {Grangier}},\ and\ \bibinfo {author} {\bibfnamefont {A.}~\bibnamefont {Browaeys}},\ }\href {https://doi.org/10.1103/PhysRevLett.104.010502} {\bibfield  {journal} {\bibinfo  {journal} {Phys. Rev. Lett.}\ }\textbf {\bibinfo {volume} {104}},\ \bibinfo {pages} {010502} (\bibinfo {year} {2010})}\BibitemShut {NoStop}%
\bibitem [{\citenamefont {Martin}\ \emph {et~al.}(2021)\citenamefont {Martin}, \citenamefont {Jau}, \citenamefont {Lee}, \citenamefont {Mitra}, \citenamefont {Deutsch},\ and\ \citenamefont {Biedermann}}]{dressing_Martin2021}%
  \BibitemOpen
  \bibfield  {author} {\bibinfo {author} {\bibfnamefont {M.~J.}\ \bibnamefont {Martin}}, \bibinfo {author} {\bibfnamefont {Y.-Y.}\ \bibnamefont {Jau}}, \bibinfo {author} {\bibfnamefont {J.}~\bibnamefont {Lee}}, \bibinfo {author} {\bibfnamefont {A.}~\bibnamefont {Mitra}}, \bibinfo {author} {\bibfnamefont {I.~H.}\ \bibnamefont {Deutsch}},\ and\ \bibinfo {author} {\bibfnamefont {G.~W.}\ \bibnamefont {Biedermann}},\ }\href {https://arxiv.org/abs/2111.14677} {\bibfield  {journal} {\bibinfo  {journal} {arXiv:2111.14677}\ } (\bibinfo {year} {2021})}\BibitemShut {NoStop}%
\bibitem [{\citenamefont {Borish}\ \emph {et~al.}(2020)\citenamefont {Borish}, \citenamefont {Marković}, \citenamefont {Hines}, \citenamefont {Rajagopal},\ and\ \citenamefont {Schleier-Smith}}]{borish_transverse-field_2020}%
  \BibitemOpen
  \bibfield  {author} {\bibinfo {author} {\bibfnamefont {V.}~\bibnamefont {Borish}}, \bibinfo {author} {\bibfnamefont {O.}~\bibnamefont {Marković}}, \bibinfo {author} {\bibfnamefont {J.}~\bibnamefont {Hines}}, \bibinfo {author} {\bibfnamefont {S.}~\bibnamefont {Rajagopal}},\ and\ \bibinfo {author} {\bibfnamefont {M.}~\bibnamefont {Schleier-Smith}},\ }\href {https://doi.org/10.1103/PhysRevLett.124.063601} {\bibfield  {journal} {\bibinfo  {journal} {Phys. Rev. Lett.}\ }\textbf {\bibinfo {volume} {124}},\ \bibinfo {pages} {063601} (\bibinfo {year} {2020})}\BibitemShut {NoStop}%
\bibitem [{\citenamefont {Jandura}\ and\ \citenamefont {Pupillo}(2022)}]{TO_Jandura2022}%
  \BibitemOpen
  \bibfield  {author} {\bibinfo {author} {\bibfnamefont {S.}~\bibnamefont {Jandura}}\ and\ \bibinfo {author} {\bibfnamefont {G.}~\bibnamefont {Pupillo}},\ }\href {https://doi.org/10.22331/q-2022-05-13-712} {\bibfield  {journal} {\bibinfo  {journal} {Quantum}\ }\textbf {\bibinfo {volume} {6}},\ \bibinfo {pages} {712} (\bibinfo {year} {2022})}\BibitemShut {NoStop}%
\bibitem [{\citenamefont {Liu}\ \emph {et~al.}(2019)\citenamefont {Liu}, \citenamefont {Hood}, \citenamefont {Yu}, \citenamefont {Zhang}, \citenamefont {Wang}, \citenamefont {Lin}, \citenamefont {Rosenband},\ and\ \citenamefont {Ni}}]{Liu2019MolecularAssembly}%
  \BibitemOpen
  \bibfield  {author} {\bibinfo {author} {\bibfnamefont {L.~R.}\ \bibnamefont {Liu}}, \bibinfo {author} {\bibfnamefont {J.~D.}\ \bibnamefont {Hood}}, \bibinfo {author} {\bibfnamefont {Y.}~\bibnamefont {Yu}}, \bibinfo {author} {\bibfnamefont {J.~T.}\ \bibnamefont {Zhang}}, \bibinfo {author} {\bibfnamefont {K.}~\bibnamefont {Wang}}, \bibinfo {author} {\bibfnamefont {Y.-W.}\ \bibnamefont {Lin}}, \bibinfo {author} {\bibfnamefont {T.}~\bibnamefont {Rosenband}},\ and\ \bibinfo {author} {\bibfnamefont {K.-K.}\ \bibnamefont {Ni}},\ }\href {https://doi.org/10.1103/PhysRevX.9.021039} {\bibfield  {journal} {\bibinfo  {journal} {Phys. Rev. X}\ }\textbf {\bibinfo {volume} {9}},\ \bibinfo {pages} {021039} (\bibinfo {year} {2019})}\BibitemShut {NoStop}%
\bibitem [{\citenamefont {Spence}\ \emph {et~al.}(2022)\citenamefont {Spence}, \citenamefont {Brooks}, \citenamefont {Ruttley}, \citenamefont {Guttridge},\ and\ \citenamefont {Cornish}}]{Spence2022GroundState}%
  \BibitemOpen
  \bibfield  {author} {\bibinfo {author} {\bibfnamefont {S.}~\bibnamefont {Spence}}, \bibinfo {author} {\bibfnamefont {R.~V.}\ \bibnamefont {Brooks}}, \bibinfo {author} {\bibfnamefont {D.~K.}\ \bibnamefont {Ruttley}}, \bibinfo {author} {\bibfnamefont {A.}~\bibnamefont {Guttridge}},\ and\ \bibinfo {author} {\bibfnamefont {S.~L.}\ \bibnamefont {Cornish}},\ }\href {https://doi.org/10.1088/1367-2630/ac95b9} {\bibfield  {journal} {\bibinfo  {journal} {New J. Phys.}\ }\textbf {\bibinfo {volume} {24}},\ \bibinfo {pages} {103022} (\bibinfo {year} {2022})}\BibitemShut {NoStop}%
\bibitem [{\citenamefont {Kaufman}\ \emph {et~al.}(2012)\citenamefont {Kaufman}, \citenamefont {Lester},\ and\ \citenamefont {Regal}}]{Kaufman2012GroundState}%
  \BibitemOpen
  \bibfield  {author} {\bibinfo {author} {\bibfnamefont {A.~M.}\ \bibnamefont {Kaufman}}, \bibinfo {author} {\bibfnamefont {B.~J.}\ \bibnamefont {Lester}},\ and\ \bibinfo {author} {\bibfnamefont {C.~A.}\ \bibnamefont {Regal}},\ }\href {https://doi.org/10.1103/PhysRevX.2.041014} {\bibfield  {journal} {\bibinfo  {journal} {Phys. Rev. X}\ }\textbf {\bibinfo {volume} {2}},\ \bibinfo {pages} {041014} (\bibinfo {year} {2012})}\BibitemShut {NoStop}%
\bibitem [{\citenamefont {Shaw}(2024)}]{ShawThesis}%
  \BibitemOpen
  \bibfield  {author} {\bibinfo {author} {\bibfnamefont {A.~L.}\ \bibnamefont {Shaw}},\ }\emph {\bibinfo {title} {Learning, verifying, and erasing errors on a chaotic and highly entangled programmable quantum simulator}},\ \href {https://simonlab.stanford.edu/theses/ShawAdam_PhDThesis.pdf} {Ph.D. thesis},\ \bibinfo  {school} {California Institute of Technology}, \bibinfo {address} {Pasadena, California} (\bibinfo {year} {2024})\BibitemShut {NoStop}%
\bibitem [{\citenamefont {Jia}\ \emph {et~al.}(2026)\citenamefont {Jia}, \citenamefont {Yu}, \citenamefont {Hu}, \citenamefont {Li}, \citenamefont {Zheng}, \citenamefont {Monroe},\ and\ \citenamefont {Covey}}]{detuning_robust_Jia2026}%
  \BibitemOpen
  \bibfield  {author} {\bibinfo {author} {\bibfnamefont {Z.}~\bibnamefont {Jia}}, \bibinfo {author} {\bibfnamefont {Y.}~\bibnamefont {Yu}}, \bibinfo {author} {\bibfnamefont {X.}~\bibnamefont {Hu}}, \bibinfo {author} {\bibfnamefont {L.}~\bibnamefont {Li}}, \bibinfo {author} {\bibfnamefont {J.}~\bibnamefont {Zheng}}, \bibinfo {author} {\bibfnamefont {C.}~\bibnamefont {Monroe}},\ and\ \bibinfo {author} {\bibfnamefont {J.~P.}\ \bibnamefont {Covey}},\ }\href {https://arxiv.org/abs/2608.26446} {\bibfield  {journal} {\bibinfo  {journal} {arXiv:2608.26446}\ } (\bibinfo {year} {2026})}\BibitemShut {NoStop}%
\bibitem [{\citenamefont {Bataille}\ \emph {et~al.}(2026)\citenamefont {Bataille}, \citenamefont {Nomura},\ and\ \citenamefont {Endres}}]{detuning_robust_Bataille2026}%
  \BibitemOpen
  \bibfield  {author} {\bibinfo {author} {\bibfnamefont {E.}~\bibnamefont {Bataille}}, \bibinfo {author} {\bibfnamefont {G.}~\bibnamefont {Nomura}},\ and\ \bibinfo {author} {\bibfnamefont {M.}~\bibnamefont {Endres}},\ }\href {https://arxiv.org/abs/2608.27426} {\bibfield  {journal} {\bibinfo  {journal} {arXiv:2608.27426}\ } (\bibinfo {year} {2026})}\BibitemShut {NoStop}%
\bibitem [{\citenamefont {Jandura}\ \emph {et~al.}(2023)\citenamefont {Jandura}, \citenamefont {Thompson},\ and\ \citenamefont {Pupillo}}]{robust_Jandura2023}%
  \BibitemOpen
  \bibfield  {author} {\bibinfo {author} {\bibfnamefont {S.}~\bibnamefont {Jandura}}, \bibinfo {author} {\bibfnamefont {J.~D.}\ \bibnamefont {Thompson}},\ and\ \bibinfo {author} {\bibfnamefont {G.}~\bibnamefont {Pupillo}},\ }\href {https://doi.org/10.1103/PRXQuantum.4.020336} {\bibfield  {journal} {\bibinfo  {journal} {PRX Quantum}\ }\textbf {\bibinfo {volume} {4}},\ \bibinfo {pages} {020336} (\bibinfo {year} {2023})}\BibitemShut {NoStop}%
\bibitem [{\citenamefont {Machu}\ \emph {et~al.}(2025)\citenamefont {Machu}, \citenamefont {Creutzer}, \citenamefont {Sayrin},\ and\ \citenamefont {Brune}}]{FoV_Machu2025}%
  \BibitemOpen
  \bibfield  {author} {\bibinfo {author} {\bibfnamefont {Y.}~\bibnamefont {Machu}}, \bibinfo {author} {\bibfnamefont {G.}~\bibnamefont {Creutzer}}, \bibinfo {author} {\bibfnamefont {C.}~\bibnamefont {Sayrin}},\ and\ \bibinfo {author} {\bibfnamefont {M.}~\bibnamefont {Brune}},\ }\href {https://arxiv.org/abs/2512.15967} {\bibfield  {journal} {\bibinfo  {journal} {arXiv:2512.15967}\ } (\bibinfo {year} {2025})}\BibitemShut {NoStop}%
\bibitem [{\citenamefont {Sheldon}\ \emph {et~al.}(2016)\citenamefont {Sheldon}, \citenamefont {Bishop}, \citenamefont {Magesan}, \citenamefont {Filipp}, \citenamefont {Chow},\ and\ \citenamefont {Gambetta}}]{1q_Sheldon2016}%
  \BibitemOpen
  \bibfield  {author} {\bibinfo {author} {\bibfnamefont {S.}~\bibnamefont {Sheldon}}, \bibinfo {author} {\bibfnamefont {L.~S.}\ \bibnamefont {Bishop}}, \bibinfo {author} {\bibfnamefont {E.}~\bibnamefont {Magesan}}, \bibinfo {author} {\bibfnamefont {S.}~\bibnamefont {Filipp}}, \bibinfo {author} {\bibfnamefont {J.~M.}\ \bibnamefont {Chow}},\ and\ \bibinfo {author} {\bibfnamefont {J.~M.}\ \bibnamefont {Gambetta}},\ }\href {https://doi.org/10.1103/PhysRevA.93.012301} {\bibfield  {journal} {\bibinfo  {journal} {Phys. Rev. A}\ }\textbf {\bibinfo {volume} {93}},\ \bibinfo {pages} {012301} (\bibinfo {year} {2016})}\BibitemShut {NoStop}%
\end{thebibliography}
\end{document}